\documentclass[twocolumn]{openjournal}
\usepackage{lipsum}

\usepackage{xcolor}
\usepackage{textgreek}
\usepackage[utf8]{inputenc}
\usepackage[english]{babel}

\usepackage{hyperref}
\hypersetup{
    unicode, 
    colorlinks=true,
    linkcolor=linkcolor,
    citecolor=linkcolor,
    filecolor=linkcolor,
    urlcolor=linkcolor,
}
\usepackage{color,colortbl}
\definecolor{linkcolor}{rgb}{0.0,0.3,0.5}
\usepackage{tensind}
\tensordelimiter{?}
\DeclareGraphicsExtensions{.bmp,.png,.jpg,.pdf}
\usepackage{verbatim}
\usepackage[normalem]{ulem}
\usepackage{orcidlink}
\usepackage{soul}
\usepackage{placeins}
\usepackage{amsmath}
\usepackage{caption}

\usepackage{subcaption}
\usepackage{xspace}
\defcitealias{des/kids:2023}{DK23}
\defcitealias{DESCSRD/etal:2018}{DESC-SRD}

\newcommand{\cosmosis}{{\tt CosmoSIS}\xspace}

\newcommand{\camb}{{\tt CAMB}\xspace}

\newcommand{\cosmolike}{{\tt CosmoLike}\xspace}
\newcommand{\polychord}{{\tt PolyChord}\xspace}

\newcommand{\augur}{{\tt Augur}\xspace}
\newcommand{\firecrown}{{\tt firecrown}\xspace}
\newcommand{\ccl}{{\tt CCL}\xspace}
\newcommand{\mcpcov}{{\tt TJPCov}\xspace}

\newcommand{\numdifftools}{{\tt numdifftools}\xspace}
\newcommand{\derivkit}{{\tt derivkit}\xspace}
\newcommand{\threept}{{3$\times$2pt}\xspace}

\newcommand{\matplotlib}{\texttt{Matplotlib}}
\newcommand{\numpy}{\texttt{NumPy}}
\newcommand{\scipy}{\texttt{SciPy}}
\newcommand{\getdist}{\texttt{GetDist}}

\def\aap{Astronomy and Astrophysics}

\def\apjl{Astrophysical Journal, Letters}

\graphicspath{ {./figures/fig_tex/} }
\usepackage{multirow}
\defcitealias{srd_paper}{SRD}
\newcommand{\dif}{\mathop{}\!\mathrm{d}}

\begin{document}
\title{Fisher Forecasting for the DESC with \augur}

\author{Paul Rogozenski\orcidlink{0000-0002-1408-6904}\altaffilmark{1,2*}}
\author{Sankarshana Srinivasan\orcidlink{0000-0003-1539-3276}\altaffilmark{3}}
\author{Javier S\'{a}nchez\orcidlink{0000-0003-3136-9532}\altaffilmark{4}}
\author{Nora Elisa Chisari\orcidlink{0000-0003-4221-6718}\altaffilmark{5,6}}
\author{Arthur Loureiro\orcidlink{0000-0002-4371-0876}\altaffilmark{7,8}}
\author{Marc Paterno\orcidlink{0000-0003-0808-8388}\altaffilmark{9}}
\author{Rebekah Polen\orcidlink{0009-0006-3437-9436}\altaffilmark{10}}
\author{Heather Prince\orcidlink{0000-0003-0028-1546}\altaffilmark{11}}
\author{Biancamaria Sersante\altaffilmark{6}}
\author{Anže Slosar\orcidlink{0000-0002-8713-3695}\altaffilmark{12}}
\author{Sandro Vitenti\orcidlink{0000-0002-4587-7178}\altaffilmark{13}}
\author{Carlos García-García\orcidlink{0000-0001-6394-7494}\altaffilmark{14,15,16,17}}
\author{Eric Gawiser\orcidlink{0000-0003-1530-8713}\altaffilmark{11}}
\author{Christos Georgiou\orcidlink{0000-0002-7950-6076}\altaffilmark{18}}
\author{C. Danielle Leonard\orcidlink{0000-0002-7810-6134}\altaffilmark{19}}
\author{Ayan Mitra\orcidlink{0000-0002-9436-8871}\altaffilmark{20}}
\author{Jeremy Neveu\orcidlink{0000-0002-6966-5946}\altaffilmark{21}}
\author{The LSST Dark Energy Science Collaboration}% \textit{Author affiliations may be found before the references.}}
\email[$^\ast$ Email:]{progozen@andrew.cmu.edu}

% \FloatBarrier % niko trying to fix floats.. not sur eif it is working
%\section*{Affiliations}
%\scriptsize
\affiliation{$^{1}$ McWilliams Center for Cosmology and Astrophysics, Department of Physics, Carnegie Mellon University, Pittsburgh, PA 15213, USA}

\affiliation{$^{2}$ Department of Physics, University of Arizona, Tucson, AZ 85721, USA}

\affiliation{$^{3}$ Universit{\"a}ts-Sternwarte, Fakult{\"a}t f{\"u}r Physik, Ludwig-Maximilians Universit{\"a}t, Scheiner-
stra{\ss}e 1, 81679 M{\"u}nchen, Germany}

\affiliation{$^{4}$ Space Telescope Science Institute, 3700 San Martin Dr, Baltimore, MD 21218, USA}

\affiliation{${^5}$ Institute for Theoretical Physics, Utrecht University, Princetonplein 5, 3584 CC, Utrecht, the Netherlands}

\affiliation{${^6}$ Leiden Observatory, Leiden University, P.O. Box 9513, 2300 RA Leiden, the Netherlands}

\affiliation{${^7}$ Oskar Klein Centre for Cosmoparticle Physics, Department of Physics, Stockholm University, Stockholm, SE-106 91, Sweden}

\affiliation{${^8}$ Astrophysics Group, Blackett Laboratory, Imperial College London, London SW7 2AZ, UK}

\affiliation{${^9}$ Fermi National Accelerator Laboratory, P.O.\ Box 500, Batavia, IL 60510-5011, USA}

\affiliation{$^{10}$ Department of Physics, Duke University, Durham NC 27708, USA}

\affiliation{$^{11}$ Department of Physics and Astronomy, Rutgers University, Piscataway, NJ 08854, USA}

\affiliation{${^{12}}$ Physics Department, Brookhaven National Laboratory, Upton NY 11973}

\affiliation{$^{13}$ Universidade Estadual de Londrina, Londrina 86051-990, PR, Brazil}

\affiliation{$^{14}$ Astrophysics, University of Oxford, DWB, Keble Road, Oxford OX1 3RH, United Kingdom}

\affiliation{$^{15}$ Waterloo Centre for Astrophysics, University of Waterloo, Waterloo, ON N2L 3G1, Canada}

\affiliation{$^{16}$ Department of Physics and Astronomy, University of Waterloo, Waterloo, ON N2L 3G1, Canada}

\affiliation{$^{17}$ CIEMAT, Avenida Complutense 40, E-28040 Madrid, Spain}

\affiliation{$^{18}$ Institut de Física d’Altes Energies (IFAE), The Barcelona Institute of Science and Technology, Campus UAB, 08193 Bellaterra (Barcelona), Spain}

\affiliation{$^{19}$ School of Mathematics, Statistics and Physics, Newcastle University, Herschel Building, NE1 7RU Newcastle-upon-Tyne, UK}

\affiliation{$^{20}$ National Center for Supercomputing Applications, University of Illinois at Urbana-Champaign 1205 W. Clark St., MC-257 Room 1008 Urbana, IL 61801, USA}

\affiliation{$^{21}$ Universit\'e Paris-Saclay, CNRS, IJCLab, 91405, Orsay, France}

\begin{abstract}
The Vera C. Rubin Observatory Legacy Survey of Space and Time (LSST) has begun its ten-year survey of the entire visible southern hemisphere. To ensure robust cosmological measurements, computationally inexpensive investigations of modeling choices must be made to gauge the performance of proposed cosmological analyses. In this paper, we introduce the \augur tool of the Dark Energy Science Collaboration (DESC), which provides Fisher forecasts for cosmological inference for the LSST using software frameworks designed for DESC science. We test the pipeline by comparing it to forecasts produced by external code and direct sampling of the posterior via nested sampling methods, finding good agreement between all methods. We additionally investigate a range of modeling and hyperparameter choices for a \threept investigation in harmonic space, providing users with diagnostics to obtain reliable forecasts. \augur will be continually updated to be compatible with the other tools in the DESC software ecosystem as additional probes and functionality become available.

\end{abstract}

% Write your keywords here

\maketitle
% {\small \tableofcontents}
% \hfill \break

%\textbf{RULE 1: ONE SENTENCE PER LINE!}

\newcommand{\PR}[1]{{\textcolor{red}{PR: #1}}}
\newcommand{\niko}[1]{{\textcolor{orange}{niko: #1}}}
\newcommand{\javi}[1]{{\textcolor{cyan}{javi: #1}}}
\newcommand{\elisa}[1]{{\textcolor{magenta}{elisa: #1}}}
\newcommand{\shankar}[1]{{\textcolor{blue}{shankar: #1}}}

\section{Introduction}
\label{sec:introduction}
The elusive nature of the accelerating expansion of the universe has puzzled cosmologists for nearly three decades.
Following the discovery of the universe’s recent accelerating expansion history \citep{reiss_1998, Perlmutter_1999}, a slew of experiments and surveys were undertaken to uncover more about the expansion history and nature of dark energy \citep{albrecht2006reportdarkenergytask}.
The completion of current-generation cosmological optical surveys is ushering in a next-generation that will, with unprecedented precision, constrain the standard cosmological model.

The next stage of wide cosmological weak lensing and large-scale structure surveys is set; the Dark Energy Spectroscopic Instrument \citep[DESI][]{DESI_Collaboration_2022} entered the stage first, completing its planned five-year data collection phase and presenting cosmological analyses of its first two years of data. The Euclid satellite \citep{euclid_2025} has already delivered a preliminary data release \citep{euclidq1}, while the NSF-DOE Vera C. Rubin Observatory has recently commenced the Legacy Survey of Space and Time (LSST). 
The timing of such Stage-IV surveys is opportune, as tensions \citep{Sch_neberg_2022, Di_Valentino_2025, Pantos_2026} between the cosmological constraints of early- and late-time observables have begun to arise, possibly pointing towards a breakdown of the concordance $\Lambda$CDM model and hinting towards the nature of the dark components that make up to $95\%$ of the energy density of the Universe today.
Further preferences of time-dependent dark energy and massless neutrinos challenge the standard model altogether \citep{desi_bao_dr1, desi_bao_dr2}.

To determine the sensitivity of upcoming data sets to the cosmological model and to establish fiducial modeling choices for cosmological analysis, forecasting capability is key. Such forecasts can be performed by exploring the likelihood function of the parameter space.
The complexity of the likelihood function typically requires numerical sampling methods, such as Markov Chain Monte Carlo (MCMC) or nested sampling, to explore the parameter space and determine constraints from (actual or synthetic) data.
Judging the constraining power of a future survey, though, need not be as computationally difficult as a full MCMC analysis.
Rather, simple estimates of the constraining power and the sensitivity to modeling choices can be performed by taking numerical derivatives of the log-likelihood to obtain a Fisher forecast \citep[][]{1997PhRvL..79.3806T}.

The expected cosmological constraining power obtained by Fisher analyses aids in determining useful degeneracy-breaking directions of data combinations, the (optimistic) constraining power of different data combinations, and provides the broader community metrics with which to judge the scientific gain from the analysis \citep[e.g., the Dark Energy Figure of Merit (DEFOM)][]{albrecht2006reportdarkenergytask, albrecht2009}.
It is important, however, to contextualize the usefulness of Fisher analyses. 
While Fisher forecasts have been successfully validated against MCMC analyses in many cases, they rely on the assumption that the log-posterior is locally well approximated by a quadratic function of the model parameters. When the posterior exhibits significant non-Gaussianity, the Fisher matrix may no longer provide reliable parameter forecasts \citep{Wolz_2012, 2016MNRAS.456L.132S}.
Separately, many large-scale structure analyses adopt Gaussian likelihood approximations for two-point statistics over the scales considered, an assumption that is often sufficiently accurate for current cosmological analyses. Under these conditions, Fisher forecasts frequently provide a good approximation to full posterior sampling results when the posterior is close to Gaussian in parameter space.
As a result, extensive Fisher analyses have been carried out by the Euclid consortium \citep{euclid_2020_forecast} for their spectroscopic and photometric probes, and the Simons Observatory \citep{Ade_2019_SO_forecast} for the majority of their analyses (except for their large-scale $B$-modes extraction), and even for the next generation of spectroscopic surveys \citep{10.1093/mnras/stad611}. 
In the past, the LSST Dark Energy Science Collaboration (LSST DESC) carried out simulated Fisher forecasts and MCMC analyses to determine the requirements for knowledge of systematic effects in the LSST DESC Science Requirements Document \citep[SRD][]{srd_paper}. 
Here, we go beyond that work to present a publicly available forecasting library that enables quick exploration of model spaces of interest to Stage-IV optical surveys.

\augur\footnote{\url{https://github.com/LSSTDESC/augur}} is a Fisher forecasting software that allows users to interface to the LSST DESC cosmology software stack, including theoretical predictions with the \texttt{Core Cosmology Library} (\ccl) \footnote{\url{https://github.com/LSSTDESC/CCL}} \citep{Chisari_2019_CCL}, analytic covariance estimation with \mcpcov \footnote{\url{https://github.com/LSSTDESC/TJPCov}}, and a generalizable likelihood evaluation framework \firecrown \footnote{\url{https://github.com/LSSTDESC/firecrown}}, allowing Fisher forecasts to be performed within the same software ecosystem used for the actual data analysis.

In this sense, the goal of this paper is not to provide updated cosmological forecasts, but rather to introduce a DESC-developed forecasting framework and establish how stable and reliable Fisher forecasts can be produced using DESC infrastructure. The work is therefore intended as a methods and validation study centered on \augur, to demonstrate its reliability and document best practices for its use. Our analysis is designed to maximize realism given the capabilities of DESC tools, while documenting the impact of the assumptions made and ensuring reproducibility.

We test \augur's capabilities using realistic analysis choices for cosmological constraints derived from the combination of galaxy clustering, cosmic shear, and galaxy-galaxy lensing (`\threept' analyses). We pay particular attention to the stability of numerical derivatives of theoretical predictions,  which are key in obtaining reliable Fisher forecasts \citep{2021arXiv210100298B}. We compare the Fisher predictions to other established codes, including \cosmosis with \firecrown, the DESC SRD Fisher pipeline, as well as nested sampling of the likelihood, to determine the validity of our assumptions as well as to find a working region of the numerical parameter choices (such as derivative step size, etc.) of a Fisher analysis for cosmology with \threept statistics in Fourier space.
The Fisher formalism also provides a framework for estimating biases arising from missing systematics \citep{Huterer_Takada_2005, Taylor_2007, 2008MNRAS.391..228A}, which can also be obtained from \augur. For those, we also investigate their region of validity. This approach has subsequently been applied to quantify systematic-induced parameter biases for Stage-IV surveys, including Euclid-like galaxy clustering and weak-lensing analyses \citep[e.g.,][]{Tutusaus_2020, Deshpande_2021, Asgar_2021_bias, euclid_2024_fisher_bias}.

This paper is structured as follows. In Section \ref{sec:fisher}, we review the Fisher Information formalism and summarize the capabilities of \augur. Section \ref{sec:methods} introduces the observables and modeling used in this work, including the definition of the two-point angular correlation function, and the model of the \threept introduced in the DESC SRD. We outline our Fisher stability tests in Section \ref{sec:tests} and report their results in Section \ref{sec:results_disscusion}, culminating in our conclusions in Section \ref{sec:conclusion}.
\section{Fisher Formalism for experimental design optimization}
\label{sec:fisher}

In this section, we review the Fisher Information Matrix formalism and how this tool is used to forecast the precision expected from cosmological surveys when constraining the parameters of the assumed cosmological model. In Section \ref{sec:fisher_formalism}, we review the form of the Fisher matrix alongside common transformations and use cases.
In Section \ref{sec:augur}, we introduce the tool used for DESC Fisher forecasting: \augur.
We detail its properties, features, byproducts, and dependencies. In this paper, we will use the convention of Greek subscripts to denote model parameters, Latin subscripts to denote tomographic bins, and capital Latin superscripts to denote tracer combinations.

\subsection{Fisher Formalism}
\label{sec:fisher_formalism}
The Fisher information matrix \citep{1922RSPTA.222..309F} is widely used in cosmological analyses to forecast the statistical precision with which model parameters can be constrained.
For a data vector $\mathbf{X}$ with likelihood $\mathcal{L} (\mathbf{X}|\boldsymbol{\theta})$, where $\boldsymbol{\theta}$ represents the set of model parameters, the Fisher matrix is defined as the expectation value of the curvature of the log-likelihood,
\begin{equation}
F_{\alpha\beta} =
\left\langle - \frac{\partial^{2}\ln{\mathcal{L}}}{\partial{\theta_{\alpha}}\partial{\theta_{\beta}}}\right\rangle_{\boldsymbol{\theta}_{\rm fid}},
\label{eq:fisher_general}
\end{equation}
evaluated at fiducial parameter values $\boldsymbol{\theta}_{\rm fid}$.
The inverse Fisher matrix provides a lower bound on the covariance of unbiased estimators $\boldsymbol{\hat{\theta}}$ of parameters $\boldsymbol{\theta}$ through the Cram\'er--Rao inequality \citep{rao1945information, cramer1946},
\begin{equation}
    \mathrm{Cov}(\boldsymbol{\hat{\theta}})
\;\succeq\;
\boldsymbol{F}^{-1}(\boldsymbol{\theta}),
\end{equation}
which is approximately saturated for a Gaussian parameter space and is commonly used as an estimate of the best achievable parameter precision in forecasting applications. Throughout this section, we will denote the parameter covariance found through the Fisher formalism as $\Sigma_{\alpha \beta} = (F^{-1})_{\alpha \beta}$. 

We assume a Gaussian likelihood, allowed by the Central Limit Theorem, which ensures that the sampling distribution of summary statistics converges to a Gaussian as the number of independent modes increases \citep[see e.g.,][for a detailed discussion on the validity of a Gaussian likelihood]{Oehl_2025}. 
For a Gaussian likelihood and parameter-independent covariance matrix, the relationship between the data and its covariance may be written as follows 
\begin{equation}
\ln \mathcal{L}
\propto
-\frac{1}{2}
\left(\mathbf{X}-\mathbf{f}(\boldsymbol{\theta})\right)^{\mathrm{T}}
\mathbf{\Sigma}_\mathbf{X}^{-1}
\left(\mathbf{X}-\mathbf{f}(\boldsymbol{\theta})\right),
\label{eq:gauss_lk}
\end{equation}
where $\mathbf{f}(\boldsymbol{\theta})$ is the model prediction for the data vector and $\mathbf{\Sigma}_\mathbf{X}$ is the covariance matrix of data vector $\mathbf{X}$.
In this case, the Fisher matrix can be expressed in terms of derivatives of theoretical predictions with respect to the model parameters
\begin{equation}
F_{\alpha\beta}
=
\left(\frac{\partial \mathbf{f}(\boldsymbol{\theta})}{\partial \theta_\alpha}\right)^{\mathrm{T}}
\mathbf{\Sigma}_\mathbf{X}^{-1}
\left(\frac{\partial \mathbf{f}(\boldsymbol{\theta})}{\partial \theta_\beta}\right).
\label{eq:fisher_gaussian}
\end{equation}
This form makes clear that, within the Gaussian approximation, parameter constraints are determined by the sensitivity of the model prediction to parameter variations, weighted by the inverse covariance of the data. The accuracy of the Fisher approximation, however, depends on how well the log-posterior (or equivalently the log-likelihood in parameter space when priors are uninformative) is locally approximated by a quadratic function around the fiducial model, rather than on the likelihood being Gaussian as a function of the data. Consequently, non-Gaussian likelihoods (e.g. Poisson likelihoods for cluster counts) may still be accurately described by the Fisher formalism provided this local quadratic approximation is valid \citep[see e.g.][]{Wolz_2012, Kodwani_2019}. Note that this expression assumes that the covariance does not depend on the model parameters, which is the case in empirically-derived covariance matrices (e.g., jackknife), but not when using model covariances. However, this is often solved by using an iterative process \citep[as executed in e.g.,][]{2009A&A...502..721E, 2025A&A...703A.158W, 2026arXiv260114559D}, and/or assuming that the derivatives of the model-based covariances are much smaller than those of $\mathbf{f}$ \citep[][]{1997PhRvL..79.3806T}.

Priors are probability distributions over parameters that encode existing information or assumptions prior to conditioning on the data used in the current analysis. 
Their shape weights different regions of parameter space independently of the likelihood.
In the case of Gaussian priors, these can be incorporated by adding the inverse prior covariance matrix to the Fisher matrix,
\begin{equation}
F_{\alpha\beta}
\;\rightarrow\;
F_{\alpha\beta} + (\Sigma_{\rm prior}^{-1})_{\alpha\beta},
\end{equation}
where $\mathbf{\Sigma}_{\rm prior}$ is the prior covariance matrix.
For independent Gaussian priors with prior width $\sigma_{\theta_\alpha}$, this reduces to adding $1/\sigma^2_{\theta_\alpha}$ to the corresponding diagonal entry of the Fisher matrix.

It is often useful to consider transformations of the parameter space.
For a change of variables from $\boldsymbol{\theta}$ to $\boldsymbol{\phi}$ with Jacobian
\begin{equation}
J_{\alpha\beta}
=
\frac{\partial \theta_\alpha}{\partial \phi_\beta},
\end{equation}
the Fisher matrix transforms as 
\begin{equation}
\mathbf{F}(\boldsymbol{\phi})
=
\mathbf{J}^{\mathrm{T}} \mathbf{F}(\boldsymbol{\theta}) \mathbf{J}.
\end{equation}
This allows expressing the parameter constraints in an alternative parametrization that may be more directly related to the physical interpretation of the results or observational sensitivity.

The Fisher formalism can also be extended to estimate biases in inferred parameter values arising from model mismatch \citep{Huterer_Takada_2005, Taylor_2007, 2008MNRAS.391..228A, BiancasThesis}.
If the model used in the analysis $\tilde{\mathbf{f}}(\boldsymbol{\theta})$ differs from the true, underlying model that generated the data $\mathbf{f}(\boldsymbol{\theta'})$, the inferred parameters are systematically shifted, where $\boldsymbol{\theta'}$ need not be the same assumed parameter space of the analysis model.
Within the Fisher approximation, the parameter bias can be estimated as
\begin{equation}
\Delta \theta_\alpha
=
(F^{-1})_{\alpha\beta}\,\tilde{B}_\beta,
\end{equation}
where the bias vector is
\begin{equation}
\tilde{B}_\beta
=
\left[
\mathbf{f}(\boldsymbol{\theta'}_{\rm true})
-
\tilde{\mathbf{f}}(\boldsymbol{\theta}_{\rm fid})
\right]^{\mathrm{T}}
\mathbf{\Sigma}_\mathbf{X}^{-1}
\left(
\frac{\partial \tilde{\mathbf{f}}(\boldsymbol{\theta})}{\partial \theta_\beta}
\right).
\label{eq:fisher_bias}
\end{equation}
This expression shows that the parameter shift is driven by the mismatch between the true and assumed model predictions, projected onto parameter derivatives and weighted by the inverse covariance of the data.

Finally, scalar summary statistics are often used to characterize the constraining power of an experiment.
A commonly used metric is the Figure of Merit (FOM) \citep{albrecht2006reportdarkenergytask}, defined for a pair of parameters $(\theta_\alpha,\theta_\beta)$ as
\begin{equation}
\label{eq:fom}
\mathrm{FOM}(\theta_\alpha,\theta_\beta)
=
\left[
\det
\begin{pmatrix}
\Sigma_{\alpha\alpha} & \Sigma_{\alpha\beta} \\
\Sigma_{\beta\alpha} & \Sigma_{\beta\beta}
\end{pmatrix}
\right]^{-1/2},
\end{equation}
where $\Sigma_{\alpha\beta}$ is the parameter covariance matrix.
This quantity is proportional to the inverse area of the confidence region in the chosen two-parameter subspace and it provides a compact measure of constraining power \citep{Hu_Jain_2004}.

\subsection{\augur}
\label{sec:augur}

In this section, we briefly introduce \augur, the LSST DESC forecasting tool, and some of its functionalities and dependencies. It is a Python API with three main components: the construction of model predictions for the data vector, a likelihood model, and an analytic covariance matrix for a given experimental setup; the computation of Fisher matrices by evaluating numerical derivatives of the model predictions with respect to cosmological and systematic parameters; and post-processing utilities to produce plots and summary metrics. In practice, \augur\ computes derivatives of the model predictions around a fiducial parameter point, assuming a parameter-independent covariance matrix, to construct the Fisher matrix for a user-defined set of parameters. The experimental setup can be specified either directly via a likelihood object or via a configuration file that describes the details of the cosmological analysis.

\augur relies heavily on \firecrown, the LSST DESC official likelihood tool implemented in Python. \augur utilizes other DESC tools, such as \mcpcov to create analytical predictions for the Gaussian covariance and \ccl to generate theoretical predictions for cosmological observables.
To summarize briefly, \firecrown creates a likelihood, $\mathcal{L}(\mathbf{X}| \boldsymbol{\theta})$, given a set of tracers (galaxy number of counts, galaxy shear, cluster number of counts, etc.), a set of statistics (two-point statistics of galaxy density, or galaxy shear, or cross-correlations), a covariance matrix, and a set of model parameters. This likelihood is a Python object and can be evaluated for different values of the model parameters, $\boldsymbol{\theta}$. \firecrown interfaces with a variety of cosmological models implemented in \ccl, as well as models for observational and astrophysical systematic effects (e.g., uncertainties in the redshift distribution of tracers, intrinsic alignments), by making the appropriate calls to \ccl. We refer the reader to the \firecrown documentation for further details\footnote{\url{https://firecrown.readthedocs.io/}}.

\augur provides additional functionalities that are useful for iterative analyses, such as applying Gaussian priors to the Fisher matrix, calculating the Fisher bias of updated parameter values, and performing parameter transformations of the Fisher matrix\footnote{In particular, \augur can transform $\Omega_c \rightarrow \Omega_m$ and $\sigma_8 \rightarrow S_8$, transforming input parameters for \ccl to parameters commonly used in \threept cosmological analyses.}. A variety of numerical differentiation techniques are available, including the  5-point stencil method described in detail in~\citet{2021arXiv210100298B}, to calculate $\frac{\partial f}{\partial \theta_{\alpha}}$: the derivatives of the model prediction for the data vector with respect to the model parameters. In the 5-point stencil method, the partial derivative of a function $f$ with respect to a parameter $x$, evaluated at $x_{0}$, and with step size $\Delta x$ is given by: 
\begin{equation}
\begin{split}
\left. \frac{\partial f}{\partial x} \right|_{x = x_0} 
&= \frac{1}{12 \Delta x} f(x_0 - 2 \Delta x)
- \frac{8}{12 \Delta x} f(x_0 - \Delta x)\\&
% + 0 \cdot f(x_0)\\&
+ \frac{8}{12 \Delta x} f(x_0 + \Delta x)
- \frac{1}{12 \Delta x} f(x_0 + 2 \Delta x).
\label{eq:5pt_stencil}
\end{split}
\end{equation}
\augur can access additional tools for numerical differentiation like \numdifftools\footnote{\url{https://numdifftools.readthedocs.io}}, which by default computes second-order Taylor expansions about the point of reference $x_0$, and \derivkit\footnote{\url{https://derivkit.org/}} \citep{derivkit}, which has a robust suite of methods and optimization techniques for numerical differentiation. 

\section{Methods}
\label{sec:methods}
In this section, we detail the models utilized to parameterize systematic uncertainties in the theoretical \threept model. As in the DESC SRD, we assume the Planck 2015 best-fit $\Lambda$CDM cosmology \citep{2016A&A...594A..13P} and consider only linear galaxy bias and intrinsic alignments as systematics in the modeling of the harmonic-space angular power spectra, where varied parameters and parameter ranges are listed in Table \ref{tab:params}. For simplicity, we do not consider additional variations to the base cosmology, such as including massive neutrinos or modeling additional astrophysical or observational systematics. These will be presented in later work. \augur has the capability to account for these parameters, so long as they are properly modeled in the \firecrown likelihood being called. In Section \ref{sec:2pcf}, we briefly summarize the form of the harmonic-space angular power spectra being modeled in this analysis, and in Section \ref{sec:SRD}, we discuss specific modeling considerations that delineate this validation analysis from the original analysis of the DESC SRD. 

\subsection{Modeling Angular Power Spectra}
\label{sec:2pcf}
For our analyses, we consider model predictions consisting of harmonic-space angular power spectra,
$f(\boldsymbol{\theta}) = C^{AB}_{ij}(\ell,\boldsymbol{\theta})$, where $A,B$ denote the probe combination and $i,j$ label tomographic bins. We consider simple models that relate the underlying cosmological signal of galaxy clustering, cosmic shear, and their cross-correlation to leading-order astrophysical systematics, including linear galaxy bias and the Non-Linear Linear Alignment (NLA) intrinsic alignment model \citep{PhysRevD.70.063526, Bridle_2007}.
In reality, there are many more nuisance parameters to consider for a full forecast, including redshift uncertainty parameters, magnification bias, shear calibration bias, higher-order intrinsic galaxy alignments, etc. \citep[see e.g.,][and references therein]{2023PhRvD.108l3519D, 2025A&A...703A.158W, 2026arXiv260114559D}. We simplify our model to match as closely as possible the original DESC SRD analysis and to more straightforwardly isolate potential artifacts from numerical calculations, with the expectation that more complex models will be validated in future LSST DESC analyses. In order to properly account for LSST Y10 sensitivity, we expect we will need to consider additional systematics to accurately model the underlying signal (e.g., higher order galaxy bias \citep{2024JCAP...02..015N}, photometric biases \citep{Awan_2025}, scatter, outliers, spectroscopic calibration biases) with additional modeling considerations necessary depending on the cosmological scales analyzed \citep[e.g., multi-parameter baryonic feedback models when including non-linear scales,][]{Chisari_2019}. We do not consider systematics outside this model in our validation of \augur, to make clearer claims about the impact of different modeling choices on our summary statistics. 

The model for the two-point angular power spectra in harmonic space is defined using the comoving distance, $\chi$, scale factor $a(\chi)$ or redshift $z(\chi)$, and the normalized number density of the galaxy sample $A$ in tomographic bin $i$: $n_i^A(z)$. For our model, we may generally write the harmonic space correlation function of tracers $A$ and $B$ using the tracers' kernel functions, $q^A_i(\chi)$, at a given multipole $\ell$ as (see e.g. \cite{})
  \begin{equation}\label{eq:cell}
    C^{AB}_{ij}(\ell)=\int \dif \chi\,\frac{q^A_i(\chi) q^B_j(\chi)}{\chi^2} P\left(k = \frac{\ell+1/2}{\chi},z(\chi)\right),
  \end{equation}
which uses the non-linear matter power spectrum $P(k,z)$ and assumes the Limber approximation \citep{Limber} at wavenumber $k$ \citep[for a discussion on the validity of this approximation, see][]{Lemos_2017}. The tracers we consider model the total weak lensing signal, denoted with the superscript $\rm \gamma$, or the galaxy clustering signal, denoted with the superscript $\rm g$. We will refer to correlations of two galaxy clustering tracers as \emph{galaxy clustering}, two weak lensing tracers as \emph{cosmic shear}, and their cross-correlation as \emph{galaxy-galaxy lensing}.

We first define the ingredient for the total observed cosmic shear signal, $q^{\rm \gamma}_i(\chi)$. We begin by defining the lensing kernel, $q_i^\kappa(\chi)$, for a tomographic bin of the normalized source distribution, $n^\kappa_i(z)$ \citep[first introduced in e.g.,][]{Kaiser_1992, Jain_1997}, as 
  \begin{equation}\label{eq:lensingkernel}
    q^\kappa_i(\chi) = \frac{3 H_0^2 \Omega_m}{2c^2}\frac{\chi}{a(\chi)}\int_\chi^{\chi_h} \dif \chi' n^\kappa_i\left(z(\chi')\right) \frac{\dif z}{\dif\chi'} \frac{\chi' - \chi}{\chi'},
  \end{equation}
where $H_0$ and $\Omega_m$ are the local Hubble expansion rate and total matter energy density today, respectively, and $c$ is the speed of light. The lensing kernel is integrated over the comoving distance along the line of sight up to the comoving distance to the horizon, $\chi_h$. 
The lensing kernel is used to calculate the theoretical expectation of the sheared galaxy's image due to weak gravitational lensing, excluding any observational or astrophysical systematics. We model the intrinsic alignment of galaxies using the NLA model, which can be incorporated as a linear combination of the lensing kernel to obtain the full observed weak lensing signal:
\begin{equation}
    q^{\rm \gamma}_i(\chi) = q^\kappa_i(\chi) + q^{\rm IA}_i(\chi).
\end{equation}
$ q^{\rm IA}_i(\chi)$ is the kernel for intrinsic alignments for the NLA model, written as
  \begin{equation}\label{eq:IA}
    q^{\rm IA}_i(\chi) = - A_1 \left(\frac{1+z(\chi)}{1+z_{\rm p}}\right)^{\eta_1} \frac{\mathrm{C}_1 \Omega_{\rm m} \rho_{\rm crit}}{D\left(z(\chi)\right)} \frac{\dif z}{\dif\chi} n^\kappa_i\left(z(\chi)\right),
  \end{equation}
where $D(z)$ is the linear growth factor normalized to $D(z=0)=1$ today, $\rho_{\rm crit}$ is the critical density of the Universe today, and $C_1$ is the normalization constant of the IA signal, often taken to be the value measured by the \textsc{SuperCOSMOS} study \citep{Brown_2002}. We model the alignment strength through the amplitude parameter $A_1$ and its redshift evolution with a power-law model, where $z_{\rm p}=0.3$ is the pivot redshift and $\eta_1$ controls the redshift dependence.

We define the kernel for tracing the galaxy overdensity field, $q_i^{\rm g}(\chi)$, which uses a linear galaxy bias model. We take the linear galaxy bias, $b_i$, to be a constant in each tomographic bin and define the galaxy clustering kernel for a tomographic bin of the lens galaxy sample, $n^{\rm g}_i$, as (see \citet{Limber, Kaiser_1992} for the first derivation of the galaxy number counts kernel)
    \begin{equation}\label{eq:gkernel}
    q^{\rm g}_i(\chi) = b_i n^{\rm g}_i\left(z(\chi)\right)\frac{\dif z}{\dif\chi}\, .
  \end{equation}

Finally, as in the DESC SRD, we model the source and lens distribution with the Smail formula \citep{Smail_1994}  
   \begin{equation}\label{eq:Smail}
       n(z) \propto \left(\frac{z}{z_0}\right)^{2}\exp\left[-\left(\frac{z}{z_0}\right)^{\beta}\right]\,.
   \end{equation}
with $(z_0 , \beta) = (0.13, 0.78)$ for the LSST Year 1 source sample, $(z_0 , \beta) = (0.11, 0.68)$ for the LSST Year 10 source sample, $(z_0 , \beta) = (0.26, 0.94)$ for the LSST Year 1 lens sample and $(z_0 , \beta) = (0.28, 0.90)$ for LSST Year 10 lens sample. 

\renewcommand{\arraystretch}{1.5}
\begin{table*}[t]
    \centering
    \begin{tabular}{|c|c|c|c|}
         \hline
         Parameter & Definition & Fiducial Value & Uniform Prior Range \\ 
         \hline
         $\Omega_\mathrm{c}$ & Dark matter density parameter & 0.26642 & [0.06, 0.46] \\
         $\Omega_\mathrm{b}$ & Baryon density parameter & 0.0492 & [0.03, 0.07] \\
         $h$ & Dimensionless Hubble parameter & 0.6727 & [0.5, 0.8] \\
         $\sigma_8$ & Matter power spectrum variance at $R= 8 \,\mathrm{ Mpc}/h$ & 0.831 & [0.3, 1.2] \\
         $n_s$ & Scalar spectral index & 0.9645 & [0.85, 1.10] \\
         $w_0$ & Dark energy equation of state at current time & -1.0 & [-3.0,-0.33] \\
         $w_a$ & Dark energy equation of state evolution parameter & 0.0 & [-3.0, 3.0] \\
         $A_1$ & Intrinsic alignment amplitude & 5.92 & [-3.0, 8.0] \\
         $\eta_1$ & Intrinsic alignment redshift evolution parameter & -0.47 & [-5.0, 5.0] \\
         \hline
         Y1 $b_i$ & Linear galaxy bias & 1.56, 1.73, 1.91, 2.10, 2.29 & [0.8, 2.5]\\
         \hline
        \multirow{2}{*}{Y10 $b_i$} & \multirow{2}{*}{Linear galaxy bias}
        & 1.38, 1.45, 1.53, 1.61, 1.69, 
        & \multirow{2}{*}{[0.8, 2.5]} \\
        &  & 1.77, 1.86, 1.94, 2.03, 2.12 
        & \\
        \hline
         
    \end{tabular}
    \caption{The fiducial values and uniform prior widths used to define the step sizes of the numerical derivative method in our \threept analysis and for nested sampling.}
    \label{tab:params}
\end{table*}

We list the varied parameters used in our analysis in Table \ref{tab:params}.

The angular power spectra are further  binned into finite bandpowers. This is a common practice used in cosmological analyses to boost the overall signal for a given cosmological scale of interest and/or account for coupled modes in the total signal of the data vector. We consider the impact of bin-averaging over our model predictions and define a bin-averaged angular power spectra, $ \bar{C}_{ij}^{AB}(\ell_{\rm eff}) $. We consider only a simple top-hat filter, taking the bin-averaged quantity to be an unweighted average of all modes, as \begin{equation}
    \bar{C}_{ij}^{AB}(\ell_{\rm eff}) = \frac{\sum_{\ell_{ij, \rm min}}^{\ell_{ij,\rm max}} C_{ij}^{AB}(\ell)}{\ell_{ij, \rm max} - \ell_{ij,\rm min} + 1} 
    \label{eq:tophat}
\end{equation}
We consider the bin-averaging model of the data alongside our fiducial choice of binning (and the choice of the DESC SRD) by modeling the angular power spectra at the average $\ell$ value of a given bin. Our model defines $\ell$-bin edges by splitting the range $20 \leq \ell \leq 15000$ into 20 logarithmically-spaced bins. In practice, the theoretical prediction for each $\ell$ bin is approximated by evaluating the angular power spectra at the geometric mean of the bin boundaries when constructing the fiducial model prediction. We note that this is a simplification; the final data analysis pipeline will involve a convolution between the theoretical predictions and the bandpower window functions.

Although the definition of our $\ell$ bin edges makes use of high ell values (small scales), we do not actually include them in our analysis. Rather, modeling at these small scales (large $ \ell$) is highly uncertain and is excluded from typical analyses via scale cuts \citep[see Section \ref{subsec:pk} and e.g.,][and references therein for some discussion on these modeling uncertainties]{2018ARA&A..56..393M, 2024JCAP...02..015N}. We apply scale cuts as in the DESC SRD, which define hard cut-offs at $\ell \leq 3000$ for cosmic shear correlations and a physical scale-cut of $k \leq 0.3 \, h/\mathrm{Mpc}$ for galaxy-galaxy lensing and galaxy clustering correlations. We note that the linear galaxy bias per tomographic bin approximation likely is not sufficient over these scales, given the sensitivity of LSST and the impact of non-linear galaxy clustering at these scales. Nevertheless, we made this choice to match the DESC SRD. 
Such a scale cut scheme is efficient since non-linear structure formation impacts smaller physical scales as one moves from tomographic bins centered at low redshift to those centered at high redshift. A more detailed analysis would employ a bin-by-bin approach to determine scale cuts. 
We also utilize $n(z)$ bins that align with the choices made in the DESC SRD (see also Section \ref{sec:SRD}).

\subsection{The \threept Model of the LSST DESC Science Requirements Document (SRD)}
\label{sec:SRD}

As mentioned before, the LSST DESC Science Requirements Document (SRD) defines benchmark Fisher forecasts for LSST Year 1 (Y1) and Year 10 (Y10), against which subsequent analyses and tools are commonly validated. In this work, we focus specifically on the harmonic-space \threept forecasts (galaxy clustering, galaxy-galaxy lensing, and cosmic shear), using the DESC SRD pipeline\footnote{\url{https://github.com/CosmoLike/DESC_SRD}}, which is powered by a lite version of \cosmolike\citep[][]{Krause_2017}, as a reference point for validating the Fisher implementation in \augur. To make a fair comparison between the DESC SRD pipeline and \augur, we outline the specific modeling considerations that enabled the DESC SRD results. We also outline the modifications made to the initial pipeline that mimic the current modeling capabilities of \ccl, \firecrown, and the rest of the DESC tools ecosystem. 

The DESC SRD modeled the \threept data vector by binning the harmonic-space angular power spectra (which we refer to as angular power spectra for the rest of this work) from $20 \leq \ell \leq 15000$ over 20 logarithmically-spaced bins, analyzing scales within the scale-cut threshold of $\ell \leq 3000$ for cosmic shear correlations and $k \leq 0.3 \, h/\rm{Mpc}$ for galaxy-galaxy lensing and galaxy clustering correlations. The particular choice of $\ell$-binning permitted analysis of cross-correlations between \threept observables and galaxy clusters at small scales. The DESC SRD did not utilize any bin-averaging over each region in $\ell$ space, but rather evaluated the harmonic space $C_{ij}^{AB}(\ell)$ at the bin center in log space. We examine the potential impact of this choice in Section \ref{sec:binavg}.

 We now turn to the modeling of the cosmological signal and the associated systematics. The DESC SRD pipeline modeled the power spectrum using analytic formulae, including the Eisenstein-Hu linear power spectrum \citep{Eisenstein_1999} and the Takahashi Halofit prescription for the non-linear power spectrum model \citep{Takahashi_2012}. 
 While we compare to this analysis choice as faithfully as we can in the DESC tools ecosystem, we find notable divergences in how different non-linear power spectrum models respond to changes in the step size of numerical derivatives (as discussed in Section \ref{subsec:pk}). The reported Fisher DEFOM values of the DESC SRD incorporated uncorrelated Gaussian priors on cosmological and nuisance parameters into the Fisher matrix\footnote{Note that the DESC SRD also included an alternative analysis with correlated `Stage III priors'; we do not refer to this alternate analysis here.}. While \augur has the capability to also incorporate these priors, we do not include them in our comparisons to isolate sources of divergences and make a more faithful comparison between the Fisher implementations. We perform our comparisons using the Limber approximation.  Additionally, we simplify the intrinsic alignment model used in the original DESC SRD to remove luminosity dependence in the amplitude, as this particular model is not available in \firecrown at the time of this analysis. We instead simplify the DESC SRD Pipeline to incorporate the NLA model for a more direct comparison. We note that the DESC SRD used a different value for the normalization of the IA amplitude, $C_1 \rho_{crit}$. \ccl instead takes this value to be $C_1 \rho_{crit} \approx 0.0139$, and the DESC SRD $C_1 \rho_{crit} = 0.0134$.  Throughout the DESC SRD Pipeline, a factor corresponding to the fractional sky area is incorporated {\it a-posteriori} to the calculated Fisher matrices. We include this factor in all comparisons and results except those shown in Figure \ref{fig:chain}, and we utilize the covariance matrix used in the original DESC SRD analysis to compute the Fisher matrix for all pipelines. 

 As in the DESC SRD forecasts, the modeling choices adopted here are intentionally optimistic and are designed to provide a controlled and reproducible baseline for Fisher validation rather than a fully realistic end-to-end cosmological analysis. In particular, the scale cuts are simplified, the redshift distribution uncertainties are represented through idealized $n(z)$ models, and intrinsic alignments are modeled with a global amplitude and redshift evolution rather than independent per-tomographic-bin parameters. These assumptions reduce parameter volume and isolate the numerical behavior of the Fisher analysis, allowing us to focus on validating the forecasting framework itself. More complex nuisance parameter treatments can be incorporated within \augur, but are beyond the scope of this work.

 Finally, we revisit the choices related to the procedure followed for numerical differentiation. The step sizes chosen to evaluate numerical derivatives for the DESC SRD 5-point stencil method were not the same absolute step size for each parameter; rather, the chosen step size was a half of the 1D marginalized Gaussian prior width used for the cosmological parameters. For the nuisance parameters, the DESC SRD implementation instead adopted a step size equal to five times the corresponding one-dimensional Gaussian prior width. We do not directly match the choice of step sizes in the original DESC SRD, but perform much of our subsequent analysis using step sizes that are normalized as fractional components of wide, uninformative, uniform prior ranges used in typical \threept  cosmological analyses. This is a convenient choice that allows the use of a step size that is of the same order of magnitude as the parameters of interest while using a single scalar value in the settings. This is particularly convenient for cases where we deal with parameters that have very different orders of magnitude (e.g., $A_{s}$, and $w_{a}$). However, there is no significance in this choice and the software allows passing a single fixed-value scalar to all parameters as step size, or an array of step-sizes (when using \texttt{numdifftools}). This choice does not imply that prior-normalized steps are intrinsically more stable than steps normalized to the fiducial parameter values. Rather, it provides a convenient and parameter-independent prescription for setting the numerical scale of the derivatives. A fiducial-value normalization can be ill-defined or undesirable for parameters whose fiducial values are close to zero (e.g. \(w_a\)) or close to unphysical regions of parameter space. The absolute step size should therefore still be checked for numerical stability for the parameter set and fiducial cosmology under consideration. Augur also allows parameter-specific step sizes to be supplied when a common normalization is not appropriate. We also note that the step size chosen for a particular Fisher analysis can produce overly-optimistic results due to underlying modeling conditions, including $\ell$-bin averaging and the choice of non-linear power spectrum. 
\section{Fisher Validation and Stability Tests}
\label{sec:tests}

In this section, we outline the tests we perform to validate the Fisher matrices calculated by \augur for use in LSST DESC forecasts. 
For our robustness tests, we specifically test the stability of \threept cosmological and systematic parameter inferences for Y1 and Y10 DESC SRD-like set-ups, examining both the width and degeneracy of the inferred Fisher contours. We ensure Fisher forecasts found via \augur are stable against potential hyper-parameters and modeling choices of the analysis, including a) the derivative method of choice, b) the step size of the derivative for numerical derivative calculations, c) the use of normalized step-sizes relative to the uniform prior range to calculate numerical derivatives, d) the choice of non-linear matter power spectrum prescription, and e) the modeling of the redshift distribution. 

We choose our fiducial power spectrum model to be the Eisenstein-Hu linear power spectrum \citep{Eisenstein_1998} coupled with the Takahashi Halofit prescription \citep{Takahashi_2012} for the non-linear matter power spectrum, mirroring the LSST DESC SRD, but also consider linear power spectrum predictions through the Boltzmann code \camb \footnote{\url{https://camb.info/}} \citep{CAMB_Lewis_2011} and the \texttt{HMCode} non-linear power spectrum prescription \citep[HMCode2020][]{Mead_2021}. We then explore the step sizes of the numerical derivatives normalized relative to physically-motivated parameter boundaries. We establish stability criteria for our fiducial step sizes by ensuring the relative difference between elements of the Fisher matrices evaluated with step sizes 50\% greater and 50\% smaller does not change by more than 10\% in any given entry. More explicitly, we enforce $(F_{\alpha \beta}^{\Delta s}-F_{\alpha \beta}^{\Delta s'})/F_{\alpha \beta}^{\Delta s'} \leq 0.1 $ for a given step size $\Delta s$ and adjacent step size $\Delta s'\in \left\{\frac{1}{2}\Delta s, \frac{3}{2}\Delta s \right\}$ for any parameter combination of $\alpha$ and $\beta$. Likewise, we ensure the stability of our analysis with respect to numerical integration by examining $n(z)$ histograms with successively finer bins, enforcing the criterion that increasing how finely sampled $n(z)$ (and the resulting kernel for integration) is with respect to $z$ should not affect the relative difference between the evaluated Fisher matrices by more than 10\% for any parameter. We vary this hyperparameter jointly with the step size, increasing the precision of the $n(z)$ spline until our stability criteria are met. Other hyperparameters, such as choices of the matter power spectrum and the accuracy parameters within \ccl and Boltzmann codes, should ideally be varied jointly to find the most numerically stable solution. For the purposes of this work, we find the default \ccl hyperparameters to be sufficiently stable.
We also record the condition number of the matrix, defined as the ratio of the maximum to the minimum eigenvalue of the Fisher matrix, which provides information on the stability of matrix inversion (crucial for obtaining reliable parameter constraints). While there is no scale to generally determine the stability of inverting matrices, we compare this condition number between Fisher evaluations to gauge whether it can be used as a stability criterion. 

After establishing a numerically stable region of parameter space, we consider the agreement between \augur and constraints obtained by alternative pipelines. We establish our agreement criteria by computing the difference between the parameter correlation matrices of each method. The parameter correlation matrix is defined as
\begin{equation}
\mathrm{Corr}_{\alpha\beta} =
\frac{\Sigma_{\alpha\beta}}
{\sqrt{\Sigma_{\alpha\alpha}\Sigma_{\beta\beta}}}.
\end{equation}
 Comparing correlation matrices rather than the Fisher matrices themselves downweights potential discrepancies in contour sizes and numerical/model implementation differences across pipelines, thereby focusing on the degeneracy directions among parameters. We find the difference between the correlation matrices of method $A$ and method $B$ and adopt a threshold of $|\mathrm{Corr}^A_{\alpha\beta} - \mathrm{Corr}^B_{\alpha\beta}| \equiv \mathrm{max}|\Delta_{\alpha \beta}| \leq 0.1$ for all $\alpha,\beta$ to determine whether the correlation matrices of Fisher matrices are in agreement with one another. This choice is a practical one, that indicates comparatively large changes in the derivative error doesn't propagate through to the correlation matrix, and therefore provides a conservative test of numerical robustness.

In our subsequent studies, we will rely on two defined FOMs using Equation \ref{eq:fom}. The first FOM is the Dark Energy FOM (DEFOM) defined as FOM($w_0$, $w_a$) which uses the CPL time-evolving Dark Energy parameters $w_0$ and $w_a$ \citep{2001IJMPD..10..213C, 2003PhRvL..90i1301L}. The second FOM is the Large-Scale Structure FOM (LSSFOM) defined as FOM($\Omega_m$, $S_8$) that uses the total matter energy density parameter $\Omega_m$ and parameter $S_8 \equiv \sigma_8 \sqrt{\Omega_m/0.3}$ with $\sigma_8$ being the amplitude of matter density fluctuations on 8 Mpc/$h$ scales.
We compare DEFOM and LSSFOM\footnote{We compare the LSSFOM only when directly available, as some pipelines do not vary $S_8$ or have parameter-transformation infrastructure.} and establish the agreement threshold when the FOMs differ by no more than 20\%. Since the Figure of Merit depends on the combined effect of all Fisher matrix elements through the parameter covariance, we adopt a slightly looser agreement criterion of 20\%, allowing for the propagation of small numerical variations into the coupled parameter constraints. Note that in practice the stable regions identified later in this work exhibit substantially smaller variations than these adopted thresholds.

The first benchmarking test we perform is to match the DESC SRD forecast. As we have detailed in Section \ref{sec:SRD}, {\tt firecrown} do not yet have all of the modeling capabilities used to create the initial DESC SRD, namely luminosity-dependent intrinsic alignment models. We therefore re-run the LSST DESC SRD pipeline with a luminosity-independent intrinsic alignment model (Equation \ref{eq:IA}) to create a more genuine comparison between those results and our validation pipeline. We also compare against the Fisher forecasts as implemented in the \cosmosis\footnote{\url{https://cosmosis.readthedocs.io/en/latest/}} modular library \citep{Zuntz_2015} that call \firecrown likelihoods, providing a direct consistency test of the numerical derivative methodology used within \augur. We additionally compare our Fisher results to constraints of the noiseless fiducial data vector using the same \firecrown likelihood via \polychord nested sampling \citep{Handley_2015}. We perform nested sampling runs to compare the parameter degeneracy directions obtained from Fisher calculations with those obtained by exploring the full posterior, where our sampling convergence criteria is set by the Bayesian evidence, achieving a tolerance of $10^{-2}$.

We finally examine the functionality of our Fisher bias implementation and investigate the limitations of a Fisher bias analysis. We perform this test by computing the Fisher bias using a data vector generated with a small shift in a single parameter. We then record the calculated output bias of the same parameter. Output Fisher biases that recover a small input bias provide a simple internal consistency check, whereas large input biases that are not recovered demarcate the region of parameter space in which the Fisher bias is a poor approximation for our model.

\section{Results and Discussion}
\label{sec:results_disscusion}
In this section, we present the results of the Fisher Matrix stability and comparison tests outlined in Section \ref{sec:tests}. In Section \ref{sec:stability}, we present the results of our stability tests regarding the choice of numerical differentiation technique. Sections \ref{sec:nz} and \ref{subsec:modeling_investigations} investigate the robustness of our results to modeling choices, including redshift distribution modeling, the methodology of bin-averaging and the non-linear prescription. Section \ref{sec:pipeline_comp} validates the results obtained via \augur by comparing to results obtained through external pipelines: \cosmosis with \firecrown and the DESC SRD Pipeline. We provide tests on the internal consistency of the Fisher Bias in Section \ref{sec:fisher_bias}. 

\begin{figure*}[t]
\centering
\begin{subfigure}[t]{0.45\textwidth}
         \centering
         \includegraphics[width=\textwidth]{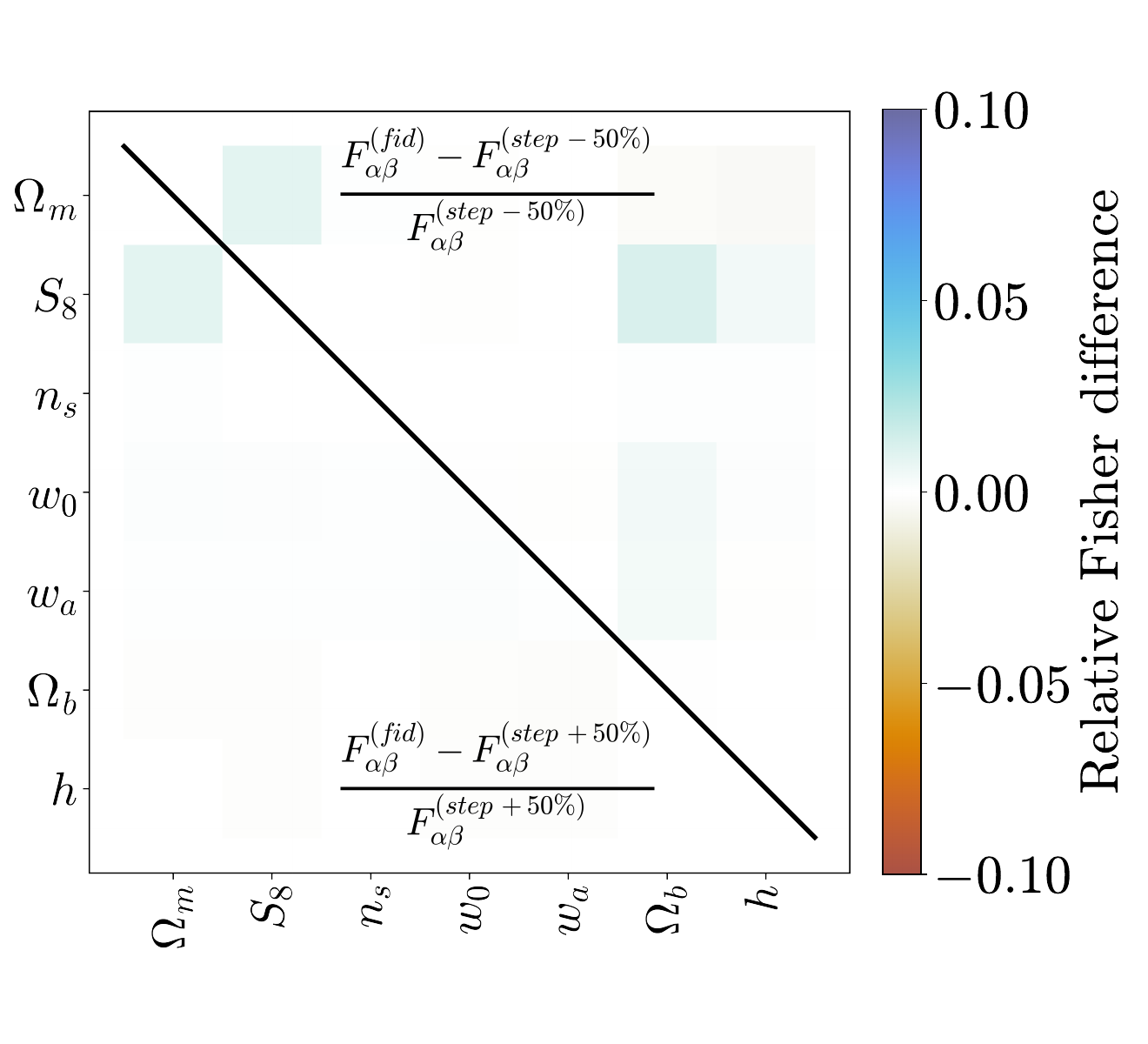}
         \caption{Relative difference in Y1 Fisher matrices computed using the fiducial normalized step size (5\% of the extent of the uniform prior range) and step sizes $50\%$ higher (lower-left triangle) and lower (upper-right triangle) than the fiducial.}
         \label{fig:y1_fisher_diff}
\end{subfigure}
\begin{subfigure}[t]{0.45\textwidth}
         \centering
         \includegraphics[width=\textwidth]{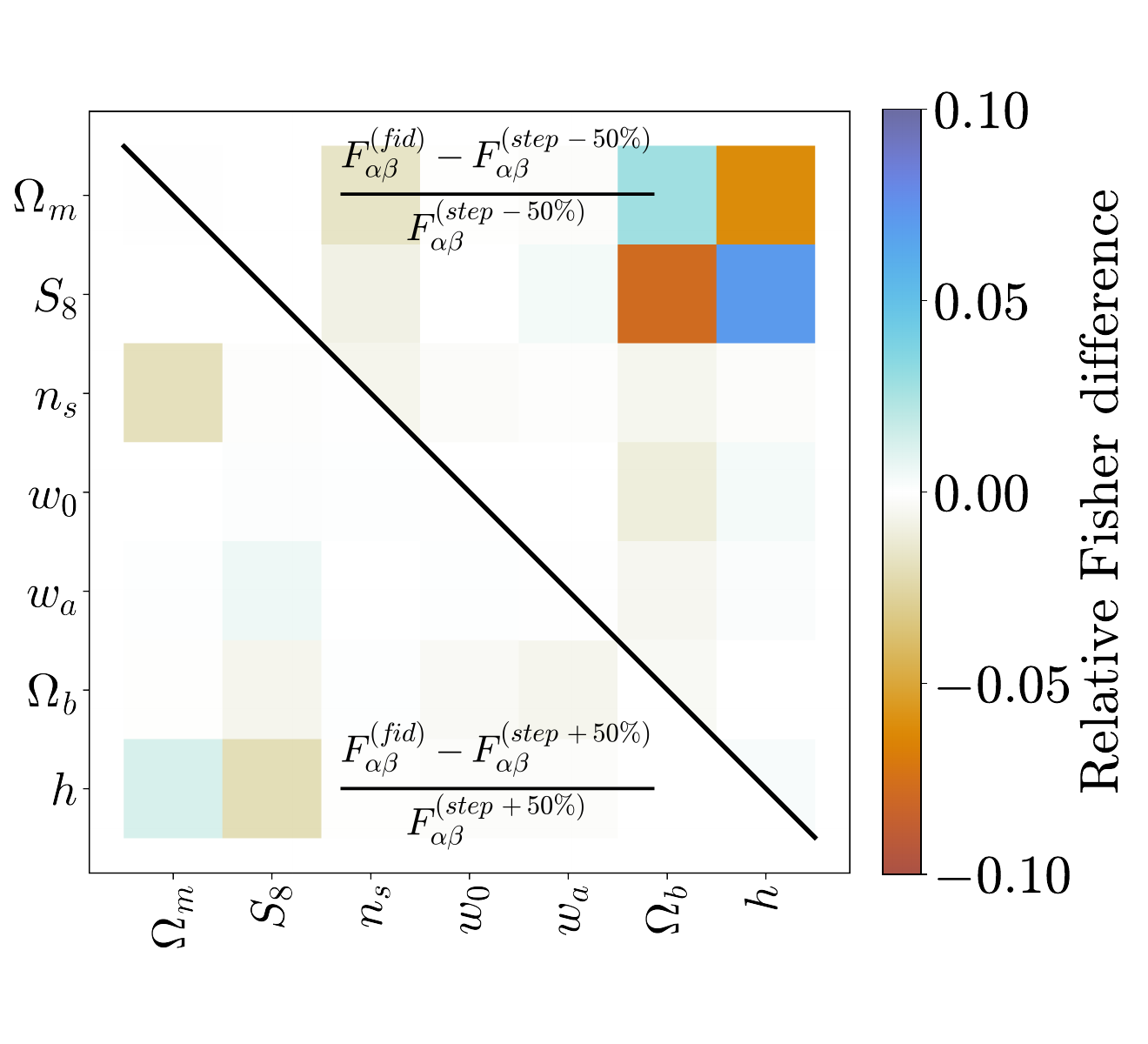}
         \caption{Relative difference in Y10 Fisher matrices computed using the fiducial normalized step size (5\% of the extent of the uniform prior range) and step sizes 50\% higher (lower-left triangle) and lower (upper-right triangle) than the fiducial.}
         \label{fig:y10_fisher_diff}
     \end{subfigure}
     
\caption{Results of the Fisher step-size stability test outlined in Sections \ref{sec:methods} and \ref{sec:tests}, which assesses the relative difference between the Fisher matrix evaluated at a proposed fiducial step size of 5\% of the extent of the uniform prior range and adjacent step sizes. The relative difference in the cosmological parameters portion of the Fisher matrices is no more than 3\% or 8\% for Y1 and Y10, respectively, confirming that our fiducial step size is a stable set-up. The results shown in this figure pertain to the 5-point stencil method, but the stability procedures confirm that this choice is a stable step size when using \numdifftools as well.}
\label{fig:fisher_diff}
\end{figure*}

\begin{figure*}[t]
\includegraphics[width=\textwidth]{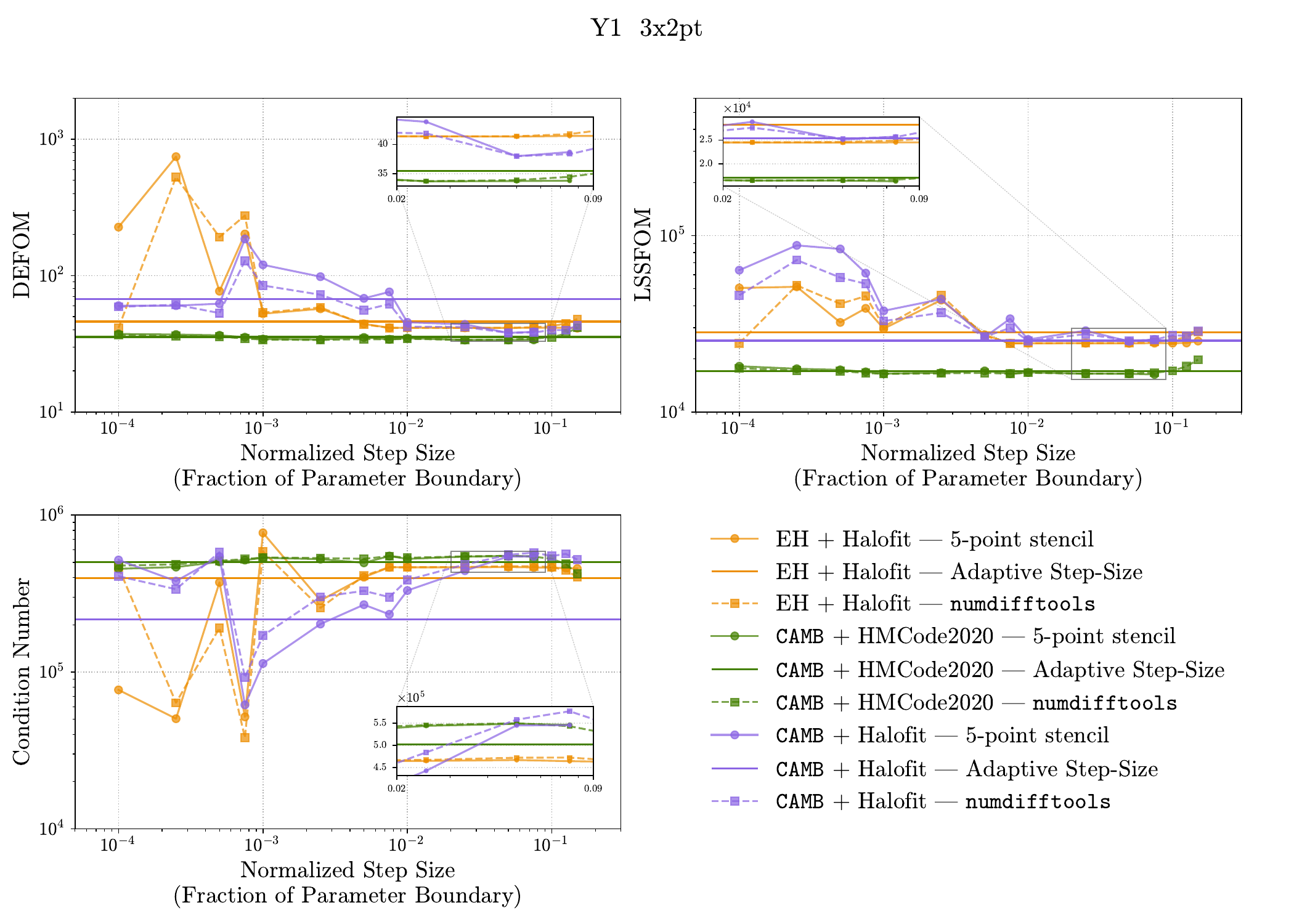}
\caption{The Y1 DEFOM (top left) LSSFOM (top right), and condition numbers (bottom left) as a function of normalized step size. To ensure robust predictions through Fisher forecasting, we explore this range of derivative step-sizes and alternative non-linear power spectrum modeling choices. Features in the non-linear power spectrum affect the numerical stability of the calculated FOMs, which are present when considering smaller step sizes about the fiducial cosmology. A common metric for assessing the stability of matrix inversion, the condition number, decreases with decreasing step size, implying improved numerical stability.}
\label{fig:y1_3x2pt}
\end{figure*}

\begin{figure*}[t]
\includegraphics[width=\textwidth]{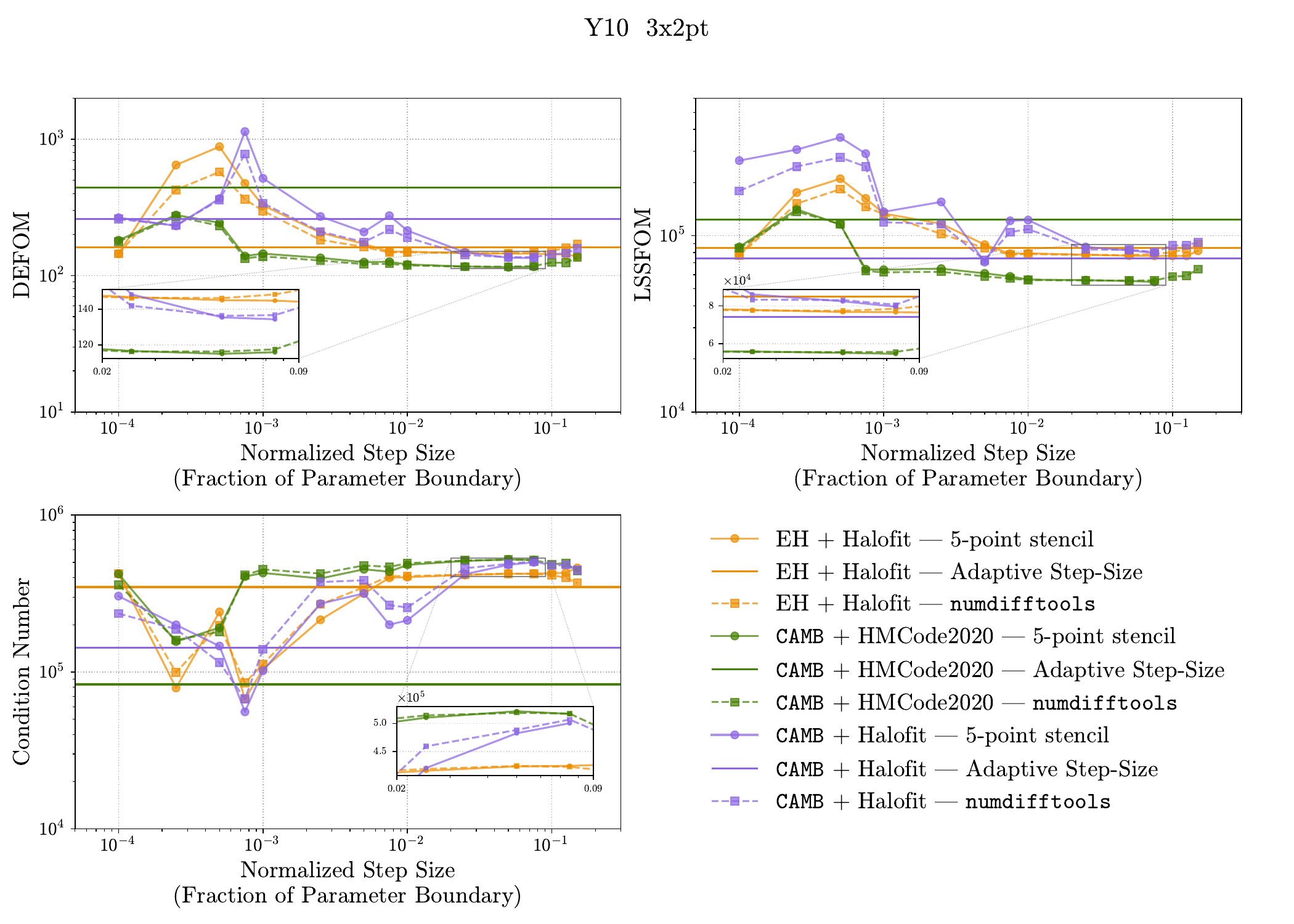}
\caption{Same as Figure \ref{fig:y1_3x2pt} but for Y10 forecasts, plotting the Y10 DEFOM (top left), LSSFOM (top right), and condition number (bottom left) with respect to normalized derivative step sizes. }
\label{fig:y10_3x2pt}  
\end{figure*}

\begin{figure*}[t]
\centering
\begin{subfigure}[t]{0.45\textwidth}
         \centering
         \includegraphics[width=\textwidth]{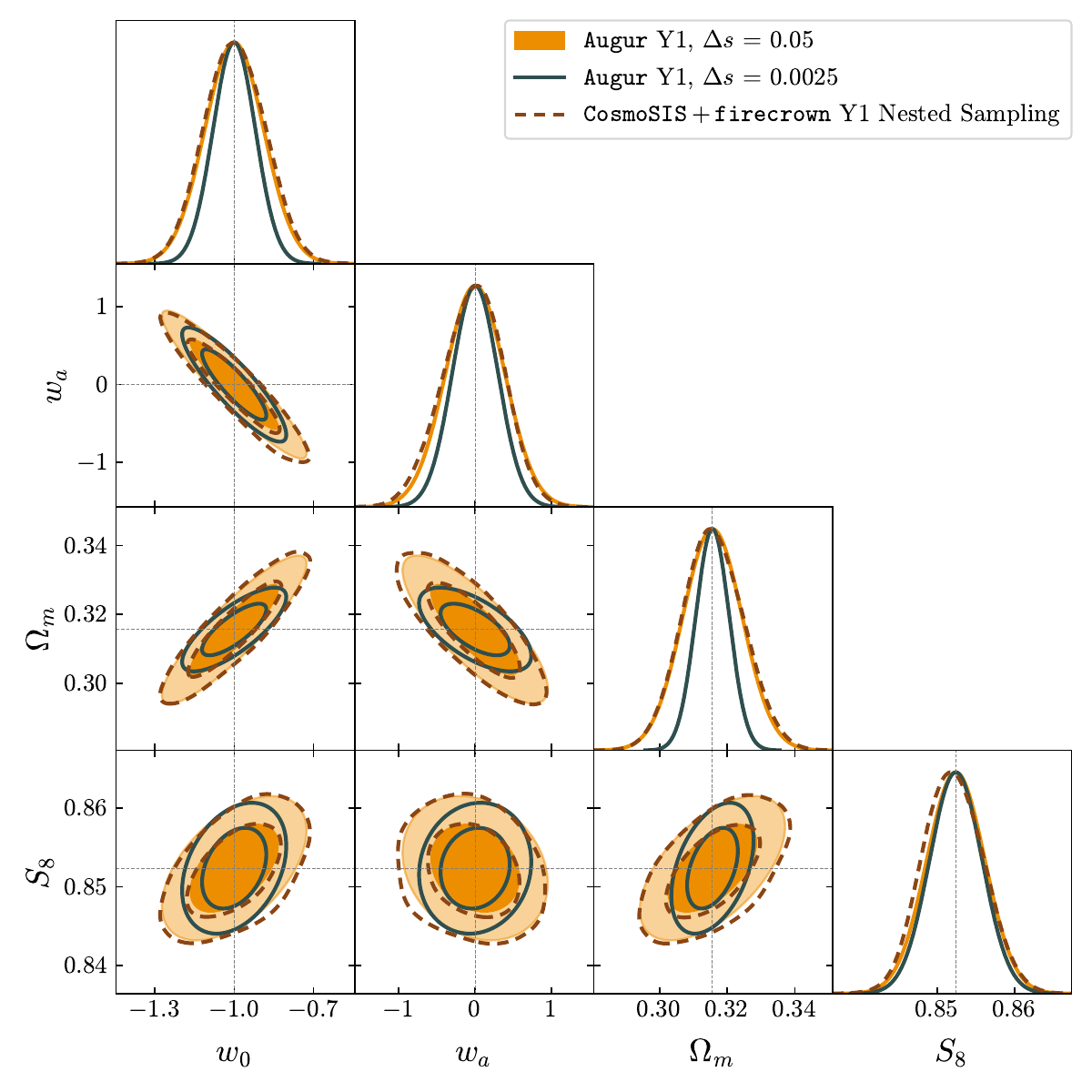}
         \caption{Forecasted Y1 Constraining Power on sampled ($w_0$, $w_a$) and derived ($\Omega_m$, $S_8$) cosmological parameters found via \augur and \polychord for our fiducial setup. The (DEFOM, LSSFOM) from the \polychord samples are (38.27, 2.29$\times 10^4$) and the \augur (DEFOM, LSSFOM) are (41.41, 2.45$\times 10^4$) at the stable step-size of $\Delta s = 0.05$. }
         \label{fig:y1_chain}
\end{subfigure}
\begin{subfigure}[t]{0.45\textwidth}
         \centering
         \includegraphics[width=\textwidth]{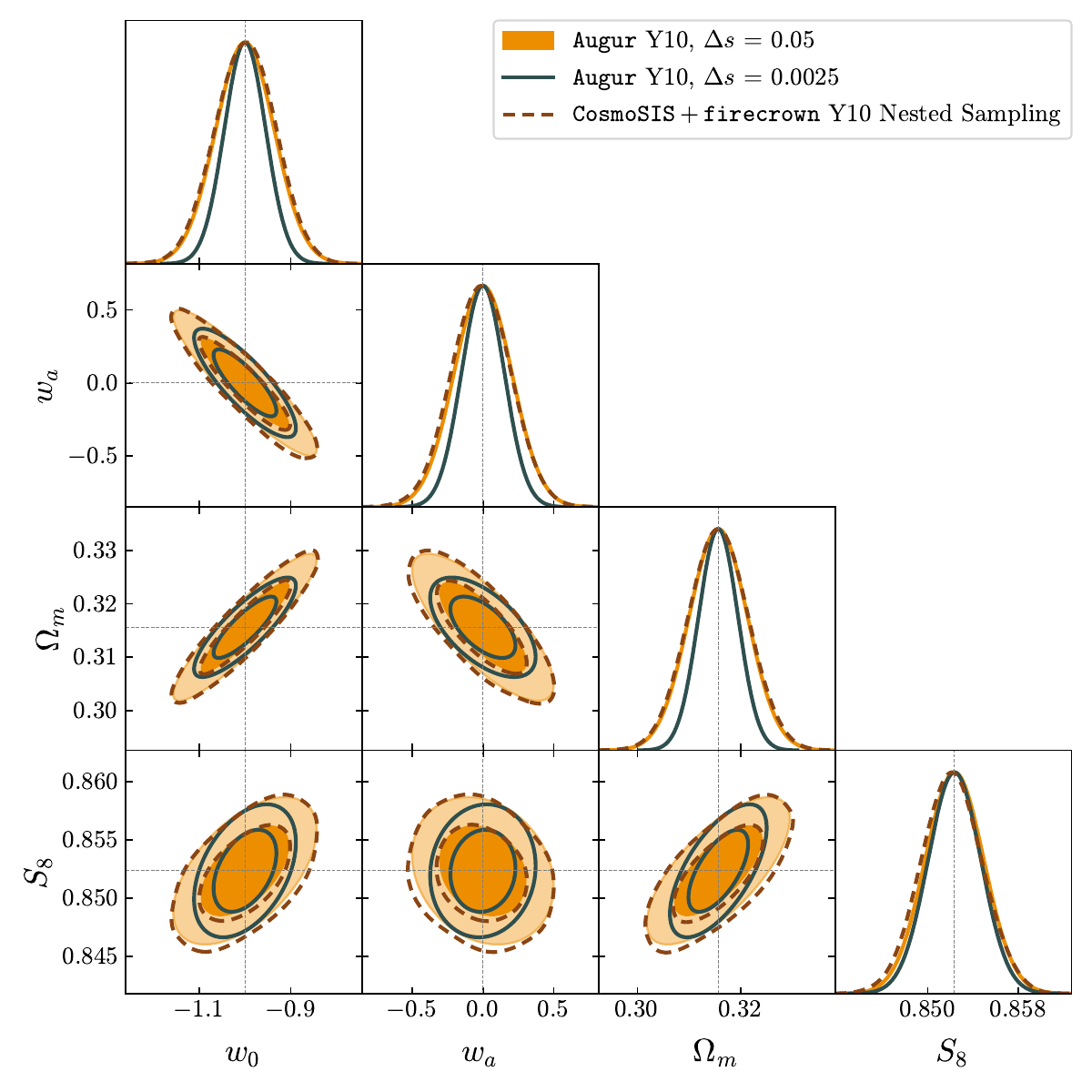}
         \caption{Forecasted Y10 Constraining Power on sampled ($w_0$, $w_a$) and derived ($\Omega_m$, $S_8$) cosmological parameters found via \augur and \polychord for our fiducial setup. The (DEFOM, LSSFOM) from the \polychord samples are (135.1, 7.1$\times 10^4$) and the \augur (DEFOM, LSSFOM) are (146.7, 7.8$\times 10^4$) at the stable step-size of $\Delta s = 0.05$.}
         \label{fig:y10_chain}
     \end{subfigure}
\caption{\augur validation tests comparing \augur Fisher method and contour widths for Y1 (Figure \ref{fig:y1_chain}) and Y10 (Figure \ref{fig:y10_chain}) to nested sampling of the likelihood through \polychord. We show the 68\% and 95\% confidence intervals of the resulting Fisher forecast and \polychord posterior samples. We transform the Fisher matrix to obtain the derived parameter $\Omega_m$ and $S_8$ and compare them to the derived parameters obtained by sampling. We plot the Fisher contours for a stable step-size ($\Delta s$ = 0.05) and one in an unstable region ($\Delta s$ = 0.0025) with the 5-point stencil method, showcasing how Fisher instability not only affects the widths but also the degeneracy direction of the parameter space. These contours also serve as validation of the Jacobian parameter transformation, as the contours of derived parameters align with those of the likelihood sampling.}
\label{fig:chain}
\end{figure*}

\begin{figure*}[t]
\centering
\begin{subfigure}[t]{0.45\textwidth}
         \centering
         \includegraphics[width=\textwidth]{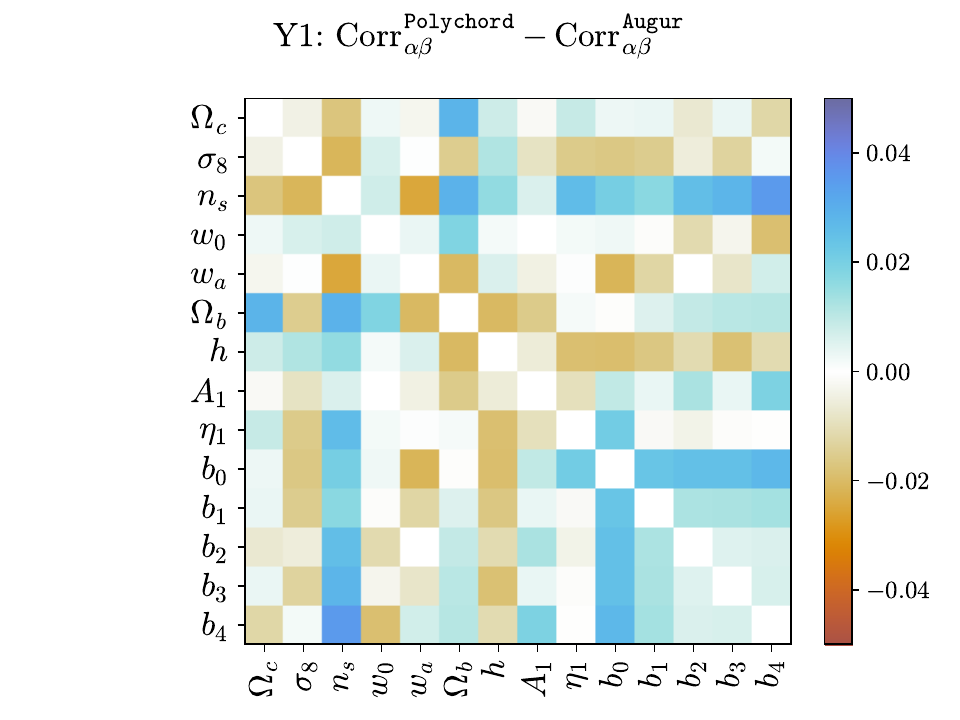}
         \caption{Difference in Correlation matrices as calculated through \augur and retrieved from nested sampling for the Y1 fiducial setup.}
         \label{fig:y1_corr_comp}
\end{subfigure}
\begin{subfigure}[t]{0.45\textwidth}
         \centering
         \includegraphics[width=\textwidth]{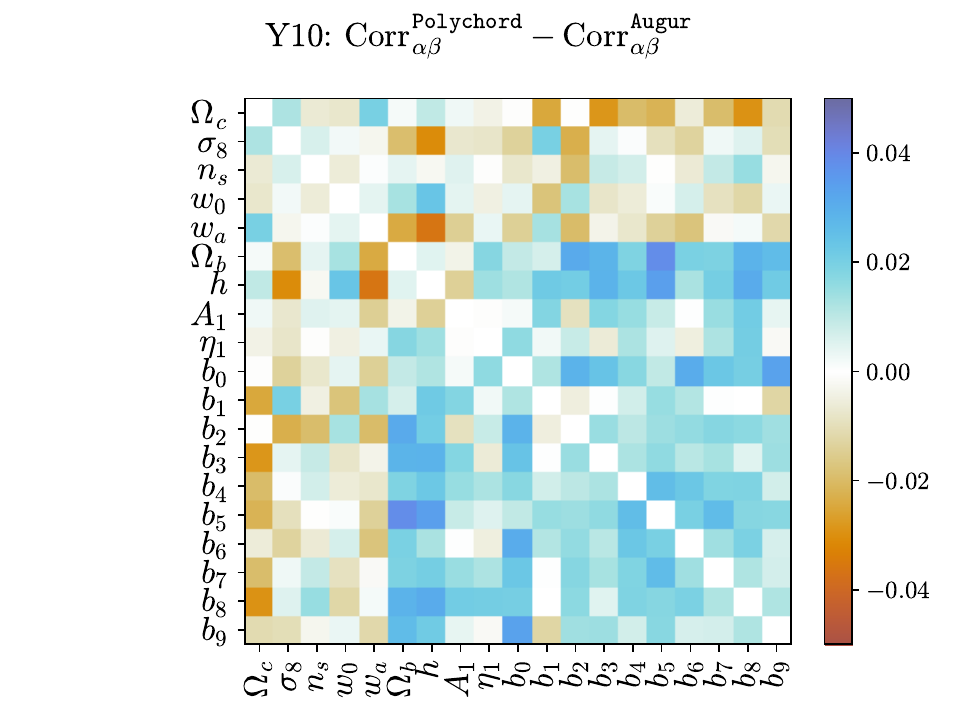}
         \caption{Difference in Correlation matrices as calculated through \augur and retrieved from nested sampling for the Y10 fiducial setup.}
         \label{fig:y10_corr_comp}
     \end{subfigure}
\caption{\augur validation tests comparing \polychord and \augur correlation matrices for Y1 (Figure \ref{fig:y1_corr_comp}) and Y10 (Figure \ref{fig:y10_corr_comp}). These tests validate the calculated derivatives of the likelihood by comparing the inferred parameter correlations to direct sampling of the likelihood. We find differences between the correlation matrices for all parameters exceed no more than 0.05, a factor of two below our proposed criteria outlined in Section \ref{sec:tests}.}
\label{fig:corr_comp}
\end{figure*}

\subsection{Fisher Matrix Stability Tests in \augur}
\label{sec:stability}

The stability and reliability of Fisher matrices depend on a variety of underlying factors, spanning from the numerical derivative technique used to the chosen model of the matter power spectrum. In this section, we investigate how some of these choices affect the stability of the calculated Fisher matrix. In the following sections, we will conduct investigations into other modeling choices.

We first determine a stable step-size normalized relative to the extent of the uniform prior range of the analysis for both Y1 and Y10, shown in Figure \ref{fig:fisher_diff}. From the procedures outlined in Sections \ref{sec:methods} and \ref{sec:tests}, we find a stable configuration for the 5-point stencil method and \numdifftools to be $\Delta s = 0.05$, or 5\% of the imposed extent of the uniform prior range. We emphasize that the prior widths are used here solely as a convenient numerical normalization for the derivative step sizes, rather than as statistical priors entering the Fisher analysis itself. Since the cosmological and nuisance parameters span several orders of magnitude, e.g., $(\Omega_{\rm b}, n_{\rm s}, A_{\rm s})$, expressing the step size as a fixed fraction of a characteristic parameter scale provides a uniform prescription that is directly comparable across parameters. The resulting normalized step size is therefore a numerical hyperparameter of the derivative calculation and should not be interpreted as coupling the Fisher forecast to the assumed prior information.

We compare the performance between the fiducial 5-point stencil method and the default settings of \numdifftools as a function of derivative step-size for LSST Y1 and Y10 in Figures \ref{fig:y1_3x2pt} and \ref{fig:y10_3x2pt}, respectively. We also compare these results to those obtained with the adaptive method implemented in \derivkit, which uses Chebyshev polynomials \citep{chebyshev1853theorie}, and plot them as a single horizontal reference line.

Between \augur's 5-point stencil method and the default \numdifftools method, we find no significant difference in the obtained FOMs for larger step sizes. As smaller step sizes are evaluated, we observe that the results obtained with either method can vary widely due to numerical uncertainties in the model. We heuristically find that starting with normalized step sizes of approximately 10\% of the prior width and decreasing until deviations between step sizes produce less than 1\% variation in the FOM yields stable results.  

As a further internal cross-check of the stable Fisher configuration identified above, we compare the resulting Fisher contours to those obtained from a full likelihood analysis using nested sampling, shown in Figure \ref{fig:chain}. For the fiducial differentiation method and step-size choice that satisfy our stability criteria, we find that the Fisher constraints closely reproduce the shape and orientation of the parameter posterior contours obtained from nested sampling. In particular, the principal degeneracy direction and the overall area of the contours are in good agreement, indicating that the linear approximation underlying the Fisher formalism is adequate in the vicinity of the fiducial cosmology for our model. 

Deviations between the Fisher and likelihood sampling constraints become more pronounced when unstable step sizes or modeling choices that lead to unstable Fisher matrices are adopted, reinforcing the importance of identifying a numerically stable regime prior to interpreting Fisher forecasts. For the default settings within \augur, the adaptive fit can occasionally lead to larger FOMs, converge to a Fisher matrix with degeneracy directions that differ from the directly sampled likelihood, or both. This comparison therefore serves not as a validation of the Fisher methodology itself, but as confirmation that \augur's implementation reproduces full-likelihood results when evaluated within its expected domain of stability.

To further quantify the agreement between the Fisher and \polychord results beyond visual contour comparisons, we compare the corresponding parameter correlation matrices, shown in Figure \ref{fig:corr_comp}. Since the Fisher formalism provides a Gaussian approximation to the posterior near the fiducial cosmology, agreement at the level of the parameter correlation structure offers a stringent test of whether \augur correctly captures parameter degeneracies. We find that, for the numerically stable configuration identified above, the Fisher and sampled correlation matrices are in close agreement, with differences well within the thresholds defined in Section \ref{sec:tests}. In particular, the dominant degeneracy between $w_0$ and $w_a$, as well as correlations between cosmological and nuisance parameters, are consistently reproduced.

This agreement at the level of both two-dimensional contours and the full correlation matrix confirms that, once evaluated in a stable numerical regime, \augur's Fisher forecasts faithfully approximate the local structure of the full likelihood. Together, these tests establish a robust baseline configuration for subsequent modeling investigations and pipeline comparisons.

\subsection{Histograms of $n(z)$}
\label{sec:nz}
Accurate knowledge of the redshift distributions of source and lens galaxies is essential for cosmological analyses based on large-scale structure observables. In Fisher forecasting studies, these distributions are often modeled using analytic forms to estimate the potential constraining power of a survey. The DESC SRD adopted such an approach, using analytic models motivated by observational expectations. In practice, however, Fisher forecasts require these distributions to be represented numerically, typically through discretized samples of the redshift distribution $n(z)$. Insufficient sampling of $n(z)$ can introduce interpolation artifacts that propagate into the theoretical predictions and their numerical derivatives. In this section, we therefore investigate how the sampling resolution of $n(z)$ affects the stability of Fisher forecasts within the \augur\ framework.

In this work, we adopt the analytic form of the DESC SRD redshift distribution, a Smail-type model \citep{Smail_1994} as in Equation~\eqref{eq:Smail}. This distribution is then divided into tomographic bins for the source and lens samples following the DESC SRD definitions.

Unlike analytic forecasting studies, \augur\ (like most cosmological inference pipelines) does not operate on continuous analytic distributions, but instead accepts $n(z)$ as a discretized set of representative knot points that are internally interpolated by \ccl\ to construct a continuous distribution. The density of these knots therefore acts as a numerical hyperparameter in the forecasting pipeline. If the distribution is sampled too sparsely, interpolation artifacts can propagate into the theoretical predictions and their numerical derivatives, while overly coarse sampling can lead to inaccurate modeling of the overlap between tomographic bins.

\begin{figure*}[t]
\centering
\begin{subfigure}[b]{0.45\textwidth}
         \centering
         \includegraphics[width=\textwidth]{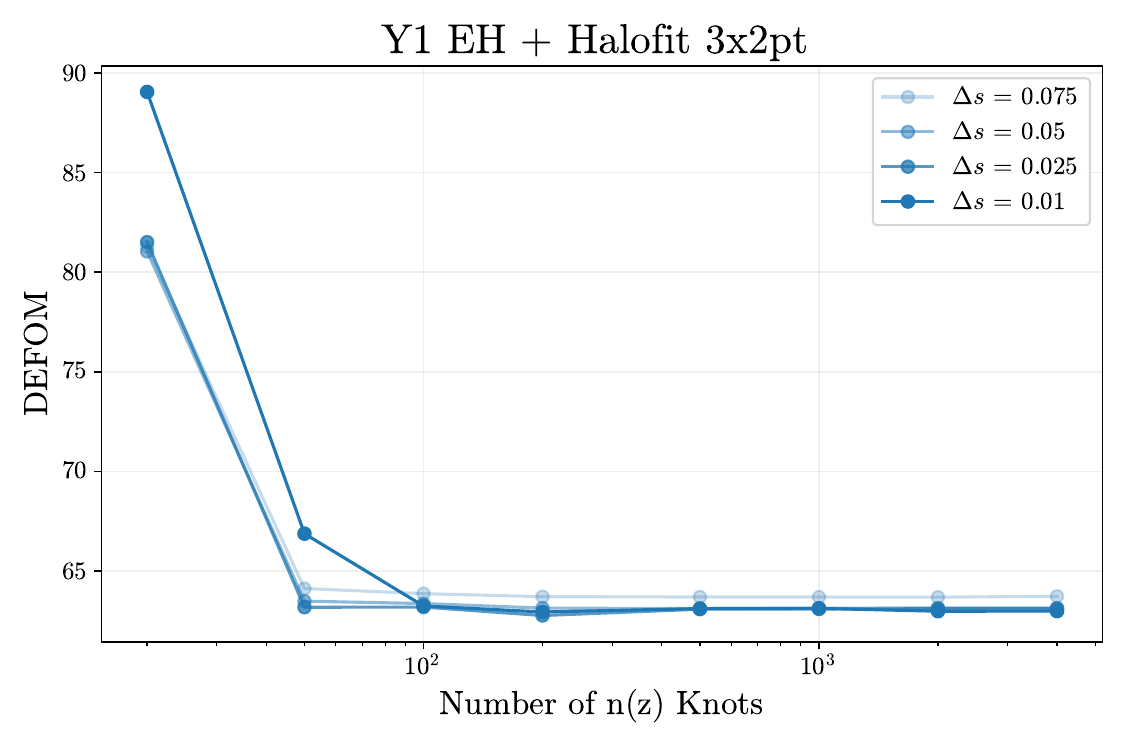}
         \caption{Variation in the Y1 DEFOM with respect to the number of knots evaluated in the $n(z)$ distribution. }
         \label{fig:y1_knots}
\end{subfigure}
\begin{subfigure}[b]{0.45\textwidth}
         \centering
         \includegraphics[width=\textwidth]{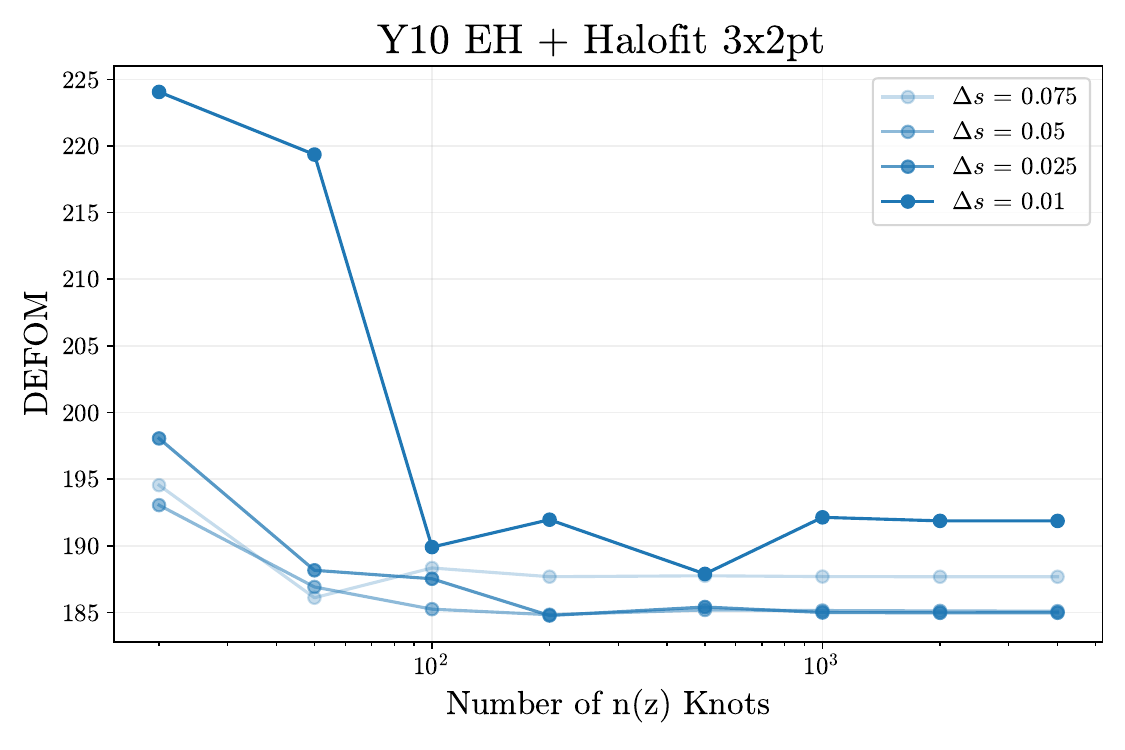}
         \caption{Variation in the Y10 DEFOM with respect to the number of knots evaluated in the $n(z)$ distribution. }
         \label{fig:y10_knots}
     \end{subfigure}
the\caption{DEFOM dependencies on the number of $n(z)$ knots for Y1 (Figure \ref{fig:y1_knots}) and Y10 (Figure \ref{fig:y10_knots}) for different derivative step-sizes, $\Delta s$. Stable solutions are generally found when increasing the number of knots in the $n(z)$ distribution to nearly 1000 evaluations, regardless of the analysis year considered.}
\label{fig:knots}
\end{figure*}
\interfootnotelinepenalty=10000
To test numerical stability with respect to the discretization of the redshift distribution, we vary the number of knots used to represent $n(z)$ and use the resulting DEFOM to portray convergence. Figure~\ref{fig:knots} shows the dependence of the DEFOM on the number of $n(z)$ knots for both Y1 (Figure~\ref{fig:y1_knots}) and Y10 (Figure~\ref{fig:y10_knots}) analyses. The tomographic bins are constructed using the \ccl examples repository (CCLX) implementation of the DESC SRD distributions\footnote{\url{https://github.com/LSSTDESC/CCLX/blob/master/LSST_SRD_Redshift_Distributions_and_Binning.ipynb}}, ensuring that the mean redshift of each bin remains stable as the number of knots is varied. This allows us to isolate the effect of sampling resolution on the predicted signal and the overlap of the tomographic tails.

As the number of knots increases, the DEFOM asymptotes to a convergent value for both Y1 and Y10 forecasts. Increasing the sampling density reduces interpolation error and provides a smoother representation of the redshift distribution, leading to stable Fisher matrices and FOM predictions. For the specific modeling configuration adopted in this work, convergence is achieved when the $n(z)$ distribution is sampled with several hundred knots. We adopt $500$ knots for our configuration, providing stable results within the step-size regime identified in Section~\ref{sec:stability}.

We emphasize that the precise sampling threshold is not universal and depends on the modeling choices adopted in a given analysis. However, this exercise demonstrates that the sampling resolution of $n(z)$ should be treated as a numerical hyperparameter in forecasting pipelines. In particular, the stability of the forecast depends on the interplay between the resolution of the input distributions and the step size used in numerical derivatives, both of which must be tested to ensure reliable Fisher results. It should also be noted that the observed distribution is often fairly noisy, and therefore, our sampling results should be treated as a somewhat `idealistic' estimate. 

While both the derivative step size and the $n(z)$ sampling density should be treated as numerical hyperparameters, our results indicate that the derivative step size is the dominant source of numerical variation once the redshift distribution is sampled with sufficient resolution. We choose 500 knots for our analysis. Beyond $\mathcal{O}(500)$ knots, further refinement of the $n(z)$ representation produces comparatively little change in the forecast for the analysis considered here.

\subsection{Additional Modeling Investigations}

\label{subsec:modeling_investigations}

Beyond purely numerical differentiation choices, modeling decisions that affect the smoothness of the model prediction as a function of $\ell$ can also influence Fisher stability. While the list of choices we investigate is by no means exhaustive, we believe that it is representative of the kind of analyses that users of \augur will be interested in carrying out. 

\subsubsection{Stability with respect to Non-Linear Matter Power Spectrum Prescription}
\label{subsec:pk}

One additional important modeling consideration is that of the non-linear power spectrum. Cosmological inference from large-scale structure observations relies on accurate and precise models of the matter power spectrum, especially beyond the regime of linear perturbation theory. A variety of non-linear matter power spectrum prescriptions have arisen over the years, from perturbation theory expansions (see \citealt{bakx2025cobraoptimalfactorizationcosmological, Bakx:2025jwa, damico2025cosmologicalanalysisdes3times2pt, DAmico:2020kxu} for recent examples) to fitting the statistics of N-body simulations (see \citealt{AbacusSummit, Hernandez-Aguayo:2022xcl} for recent examples). The most successful models for \threept observables can describe modifications to the linear power spectrum down to small scales across a variety of cosmological models with good accuracy. For this reason, practical implementations typically rely on fitting functions calibrated to simulations (see e.g., \citealt{Takahashi_2012, Mead_2015, Mead_2021, Bartlett_2024}) or machine learning-based emulators (see e.g., \citealt{Aric_2021, EE2_2021, Moran_2022, Stadler_2023}), which allow these non-linear corrections to be evaluated efficiently within cosmological inference pipelines. In this work, we explicitly consider two models for the linear matter power spectrum (Eisenstein-Hu, abbreviated to EH \citep{Eisenstein_1998, Eisenstein_1999} and \camb\footnote{\url{http://camb.info/}} \citep{CAMB_Lewis_2011, Lewis_2000}, and two for the non-linear prescription \citep[HMCode2020 and Halofit][respectively]{Mead_2021, Takahashi_2012}. While each individual implementation of the non-linear matter power spectrum can provide accurate predictions, each prescription depends uniquely on the cosmological parameters. The model's dependence on the cosmological parameters, as well as oscillatory features in $P(k)$ as a function of $k$, alters the sensitivity to step size of the numerical derivatives of the model prediction and the final prediction of the summary statistics, i.e., the $C_{ij}^{AB}(\ell)$. 

We show this effect for three commonly-used non-linear prescriptions in Figures \ref{fig:y1_3x2pt} and \ref{fig:y10_3x2pt} using the same $n(z)$ histogram found in Section \ref{sec:nz}. The first uses the same DESC SRD model, where we use the EH linear power spectrum fit with the Takahashi Halofit (Halofit) prescription, implemented in \ccl. The other two use the linear power spectrum from the \camb Boltzmann code, with either the Halofit or the HMCode2020 non-linear power spectrum model, as implemented in \camb. The Takahashi Halofit prescription expresses non-linear corrections as fitting functions of cosmological parameters. In contrast, HMCode2020 introduces these corrections through modifications to the halo model, whose parameters depend on quantities such as the matter density, growth function, and halo mass function (among others). 

The stability of the derivatives of the power spectrum model with respect to the model parameters is depicted by evaluating the DEFOM and the LSSFOM by decreasing the normalized step size of the numerical derivative and searching for convergence in the calculated values. We find that decreasing the step size below a characteristic numerical resolution does not lead to monotonic convergence of the derivatives. For sufficiently small step sizes, we recover the canonical result that numerical noise introduced by the discrete evaluation of the likelihood dominates, leading to large fluctuations in the inferred Fisher matrix elements. This is particularly true for models that utilize the Halofit non-linear prescription, which produces oscillatory features in the matter power spectrum as a function of wavenumber. These features can lead to large deviations when computing numerical derivatives with very small parameter step sizes. This feature is persistent regardless of the numerical derivative method. Additional tests use the adaptive Chebyshev polynomial fit, as implemented in \derivkit, to identify configurations of numerical derivatives that do not vary significantly across parameter step sizes. These tests reaffirm that features in the non-linear power spectrum can create configurations that are local extrema of the underlying derivatives, which often lead to unrealistically large values of the FOMs. Instead, the most stable configurations of the FOMs with respect to the normalized derivative step-size appear when taking derivative steps on the order of a few percent of the wide, uninformative priors. Interestingly, different non-linear power spectrum models exhibit varying degrees of dependence on cosmological parameters. We observe this in the Figures \ref{fig:y1_3x2pt} and \ref{fig:y10_3x2pt}, where the DEFOM and LSSFOM can change by a factor of 10\% or more in the stable region, depending on the non-linear model used. We also observe that the condition number obtained by the Fisher matrix calculation cannot be directly used as a robust indicator of reliable Fisher results. While a lower condition number indicates better numerical stability of inverting the Fisher matrix, minimizing the condition number must not be confused with determining Fisher calculation stability. In other words, the stability and minimization of the condition number is not sufficient evidence for a stable configuration, but is necessary, since large condition numbers indicating unstable Fisher matrices.

We finally compare the Fisher matrix results using the EH + Halofit power spectrum model to those obtained through \polychord nested sampling of the posterior for Y1 (Figure~\ref{fig:y1_chain}) and Y10 (Figure~\ref{fig:y10_chain}), verifying that the Fisher matrix provides a good approximation to the posterior obtained through direct sampling. The results in Figure \ref{fig:chain} display $\Omega_m$ and $S_8$ contours which are not directly varied in \augur but are the result of a transformation of variables. We compare the likelihood sampling results from \polychord to the transformed Fisher matrix obtained via that 5-point stencil method also to validate that the transformations we apply match those obtained from sampling for two different normalized step sizes. We find that the percent-level normalized step size provides good agreement in the degeneracy directions and FOMs found from sampling, whereas sub-percent step sizes generate overly optimistic results with parameter degeneracy directions discrepant from the nested sampling runs. These results are not only consistent with previous Fisher analyses, but also support our procedures for finding the best normalized step size for a Fisher analysis, which seeks the region of stability between coarse derivative step sizes and numerical artifacts enhanced by exceedingly small step sizes. 

\subsubsection{Bin-Averaging/Bandpowers}
\label{sec:binavg}

Cosmological analyses of large-scale structure do not typically evaluate single modes in harmonic space or exact angular separation scales in the sky. Instead, some binning of the data is utilized to increase the signal-to-noise of the measurement. We explore the impact of this modeling approach through simple top-hat filters in harmonic space (Equation \ref{eq:tophat}) and how the stability of the FOMs responds to bin-averaging.

\begin{figure*}[t]
\centering
\includegraphics[width=\textwidth]{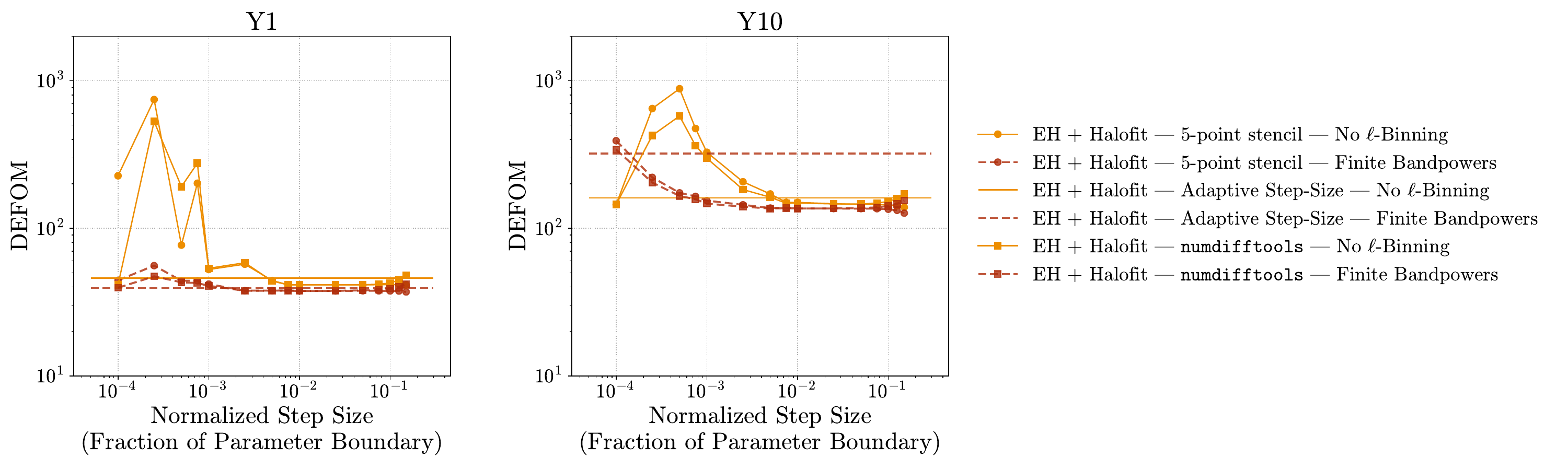}
\caption{Variation in the DEFOM when utilizing Top-Hat Bandpower Filters (Finite Bandpowers) when calculating the \text{modeled} $C(\ell)$'s for Y1 (left) and Y10 (right). We also plot the fiducial analysis that does not contain bandpowers (No $\ell$-Binning). We utilize the same covariance matrix as in the fiducial analysis, isolating potential effects of $\ell$ bandpower-averaging in the above figures. An overall lower DEFOM is calculated and smoother variations in the DEFOM with respect to the step size are observed. }
\label{fig:tophat}
\end{figure*}

Our results are shown in Figure \ref{fig:tophat} for the Y1 and Y10 DEFOM, using the identical hyperparameter setup as above. 
We find that bin-averaging smooths the variations observed in Figures \ref{fig:y1_3x2pt} and \ref{fig:y10_3x2pt} for small step sizes. 
A portion of this behavior occurs since bin-averaging effectively averages over neighboring multipoles, reducing the sensitivity of the derivatives to fluctuations at individual $\ell$ modes and mitigating numerical instabilities.
However, a larger contribution arises from the definition of our scale-cuts, which retain bin combinations only if the entire bandpower lies within the allowed region. An additional 7/15 of the allowed 335/549 data points in Y1/Y10 are excluded for some small-scale galaxy clustering and galaxy-galaxy lensing data, leading to a generally lower DEFOM relative to Figures \ref{fig:y1_3x2pt} and \ref{fig:y10_3x2pt}.

\subsection{Fisher Pipeline Comparisons}
\label{sec:pipeline_comp}

Having established internal numerical and modeling stability, we now test reproducibility across independent Fisher implementations. Explicitly, we compare the output of \augur Fisher matrix evaluations to those obtained through alternative pipelines, holding the underlying likelihood, covariance, and theoretical predictions fixed. Agreement would therefore serve as a validation of the Fisher implementation in \augur, rather than of the underlying physical model.

The first comparison we make is with results obtained using the 5-point stencil method in the \cosmosis framework, which uses the same \firecrown likelihood as \augur. This test provides a direct comparison between the numerical derivative methodologies in \augur and those in an established, external code. 
We also compare the output of \augur to evaluations obtained by the DESC SRD Pipeline. Again, \augur depends on \firecrown likelihoods powered by \ccl, whereas the DESC SRD pipeline\footnote{\url{https://github.com/CosmoLike/DESC_SRD}} uses a lightweight version of \cosmolike\footnote{\url{https://github.com/CosmoLike/cosmolike_light}} \citep{Krause_2017} to produce harmonic-space model predictions and internally compute the Fisher matrix with the 5-point stencil method. The following analyses therefore provide tests of reproducibility (i.e., that new infrastructure can reproduce or account for departures from legacy pipelines, as in Section \ref{sec:SRD}) and a moderate code comparison. We focus our comparisons on \threept data vectors for Y1 and Y10 forecasts, comparing the DEFOM and correlation matrix ratios across runs.

\begin{figure*}[t]
\centering
\begin{subfigure}[t]{0.45\textwidth}
         \centering
         \includegraphics[width=\textwidth]{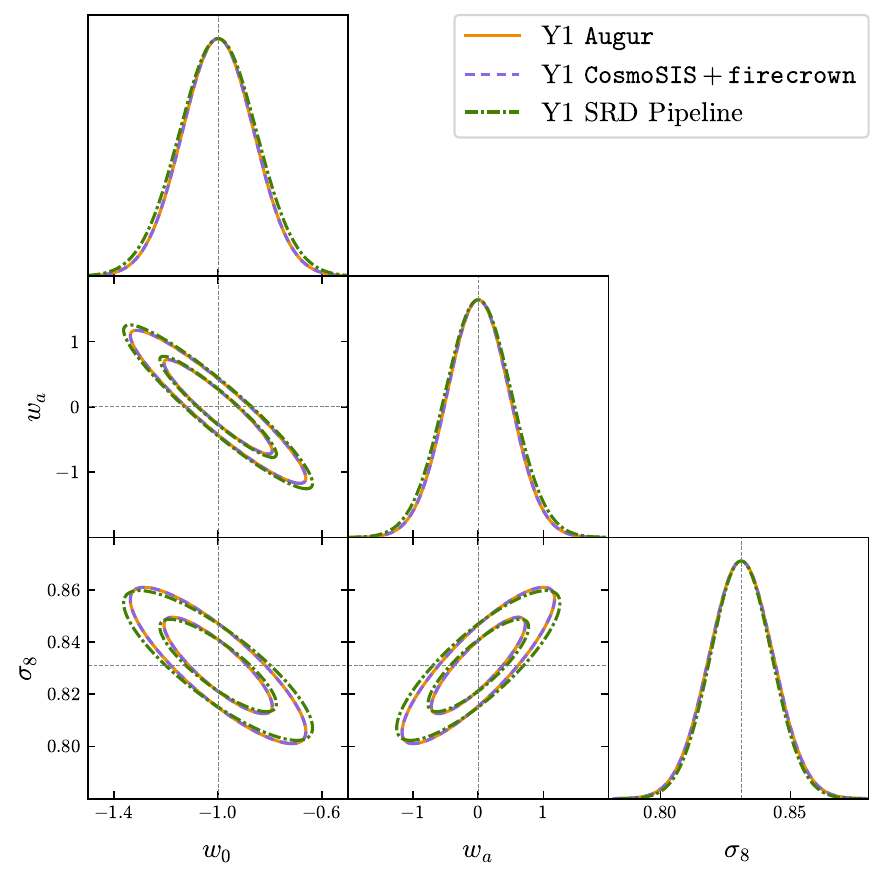}
         \caption{Forecasted Y1 Constraining Power on the $w_0$, $w_a$, and $\sigma_8$ cosmological parameters. The calculated DEFOM values are 36.95, 41.36, and 41.36, obtained via the DESC SRD Pipeline, \cosmosis with \firecrown, and \augur, respectively. }
         \label{fig:y1_triangle_comp}
\end{subfigure}
\begin{subfigure}[t]{0.45\textwidth}
         \centering
         \includegraphics[width=\textwidth]{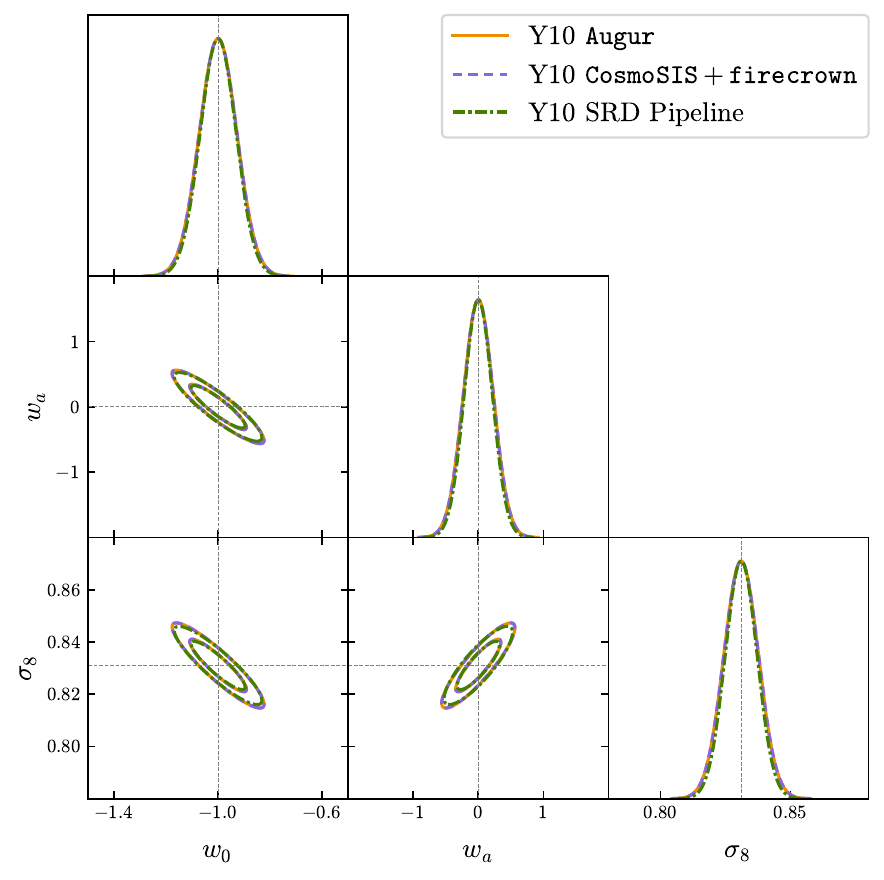}
         \caption{Forecasted Y10 Constraining Power on the $w_0$, $w_a$, and $\sigma_8$ cosmological parameters. The calculated DEFOM values are 147.2, 145.0, and 145.0, obtained via the DESC SRD Pipeline, \cosmosis with \firecrown, and \augur, respectively. }
         \label{fig:y10_triangle_comp}
     \end{subfigure}
\caption{\augur validation tests comparing \cosmosis Fisher methods and predictions using the DESC SRD Pipeline for Y1 (Figure \ref{fig:y1_triangle_comp}) and Y10 (Figure \ref{fig:y10_triangle_comp}). These tests showcase \augur's ability to reproduce the results of external cosmological forecasting pipelines and Fisher matrix computations. Each of these Fisher analyses utilizes the 5-point stencil method through their own internal mechanisms. While the \cosmosis and \augur pipelines shown use a normalized step size of $\Delta s=0.05$, we maintain the original step sizes utilized in the DESC SRD analysis to find the DESC SRD Pipeline contours. We find near-exact agreement between the \cosmosis 5-point stencil method and the \augur implementation for Y1 and Y10. We additionally find 10\% and 2\% agreement between the DEFOMs calculated between \augur and the DESC SRD Pipeline for Y1 and Y10, respectively, passing our convergence criteria outlined in Section \ref{sec:tests}.}
\label{fig:srd_contour_comp}
\end{figure*}

Our pipeline comparison results are shown in Figure \ref{fig:srd_contour_comp}. We find \augur produces results consistent with the 5-point stencil derivative method implemented within \cosmosis. Additionally, the DEFOM obtained by the DESC SRD pipeline is within 10\%/2\% consistent with the DEFOM found by \augur for Y1/Y10. Discrepancies between the DESC SRD Pipeline and \augur primarily stem from differences in the binned $n(z)$ distributions and the integration of bin combinations with large redshift separations. 

Comparing the correlation matrices of the runs, we find near exact agreement between the \cosmosis and \augur pipelines, since they both call the same underlying \firecrown likelihood. Additionally, we find all cosmological parameters pass the criteria of correlation matrix differences being $\mathrm{max}|\Delta_{\alpha\beta}| \leq 0.1$ when comparing the DESC SRD Pipeline to \augur with the exception of the $n_s$ parameter, where differences in $n_s$ correlations can reach levels of 20-23\% depending on the survey year considered. Alongside $h$ and $\Omega_b$, $n_s$ is a parameter that is weakly constrained by \threept cosmology, and discrepancies between codes can be exaggerated in weakly-constrained parameters. All correlations in the strongly constrained $\Omega_m-\sigma_8-w_0-w_a$ parameter space are within $|\Delta_{\alpha\beta}| \leq 0.03$. Critically, we find that the \augur correlation matrices are within $|\Delta_{\alpha\beta}| \leq 0.05$ of those of the sampled likelihood for all parameters, including systematic parameters, as shown in Figure \ref{fig:corr_comp}. We find the greatest discrepancies in the correlation matrices in $h$ and $\Omega_b$, as expected given that these are weakly constrained by \threept cosmology. We also compare a pure \cosmosis run with \augur predictions, using \camb + HMCode2020 as the input power spectrum model for \threept calculations, and with additional linear systematics (i.e., a multiplicative shear calibration parameter). Differences in $\ell$-interpolation and \camb settings can lead to discrepancies when comparing the relative difference in the Fisher matrices (especially for poorly constrained parameters). Despite implementation differences, we find that the differences in the correlation matrices are well within our agreement criteria of $\mathrm{max}|\Delta_{\alpha\beta}| \leq 0.1$ for both the Y1 and Y10 setups.

\subsection{Fisher Bias Internal Consistency Tests}
\label{sec:fisher_bias}

\begin{figure*}[t]
\centering
\begin{subfigure}[t]{0.45\textwidth}
         \centering
         \includegraphics[width=\textwidth]{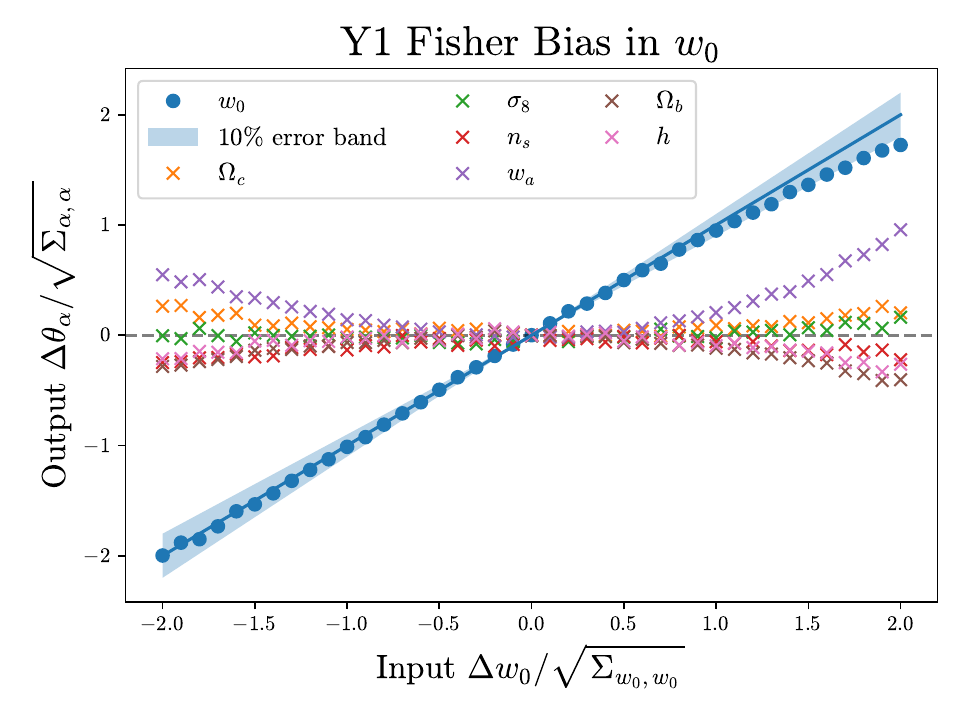}
         \caption{Fisher Bias Input and output shifts for parameter $w_0$, where shifts are normalized by the 1$\sigma$ confidence interval obtained by the Fisher matrix (in this case, $\sigma_{w_0} = 0.12$).}
         \label{fig:fisher_bias_w0}
\end{subfigure}
\begin{subfigure}[t]{0.45\textwidth}
         \centering
         \includegraphics[width=\textwidth]{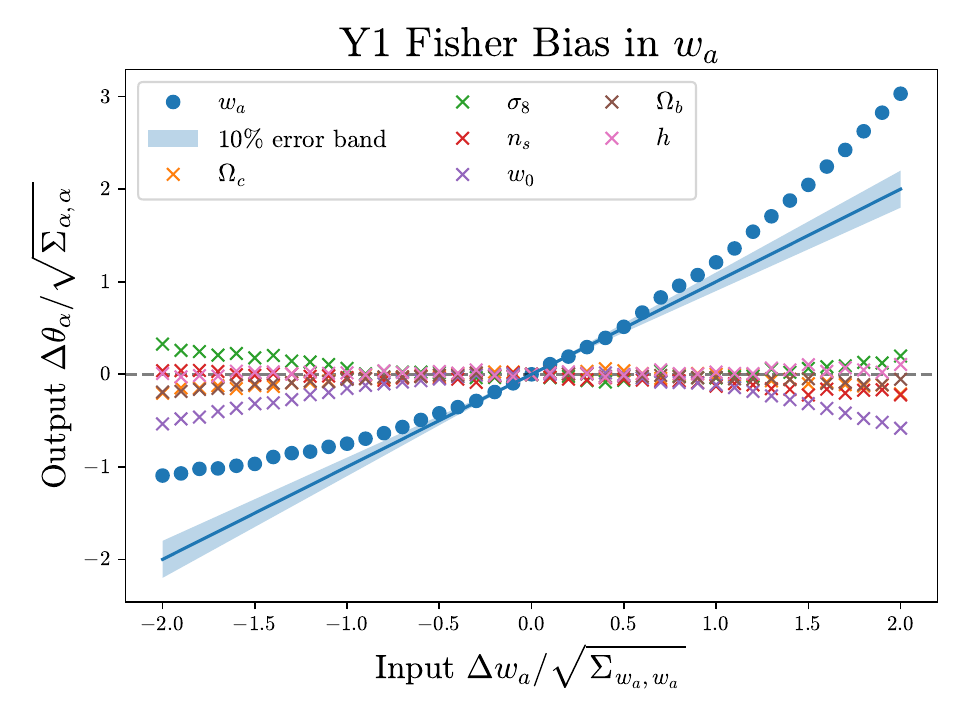}
         \caption{Fisher Bias input and output shifts for parameter $w_a$, where shifts are normalized by the 1$\sigma$ confidence interval obtained by the Fisher matrix (in this case, $\sigma_{w_a} = 0.42$).}
         \label{fig:fisher_bias_wa}
     \end{subfigure}

\caption{Fisher Bias recovery tests for the Y1 set-up. We generate model predictions biased by a single parameter relative to the fraction of the inferred contour width as input to the Fisher Bias calculation (x-axis) and assess whether the output Fisher Bias implemented in \augur recovers the input value (y-axis). We show these tests when performing a Fisher Bias on $w_0$ (Figure \ref{fig:fisher_bias_w0}) and $w_a$ (Figure \ref{fig:fisher_bias_wa}) alongside the resulting Fisher Bias shifts in the cosmological parameters. Small parameter biases about the fiducial input are recovered, and \augur produces reliable shifts in the remaining set of cosmological parameters. Large biases from the fiducial input generally break the linear assumptions of the Fisher Bias formalism and are not robustly recovered.}
\label{fig:fisher_bias}
\end{figure*}
 
The results of the Fisher Bias consistency tests are presented in Figure \ref{fig:fisher_bias}. In this exercise, we impose a controlled one-dimensional offset on a cosmological parameter ($w_0$ or $w_a$), and evaluate the bias predicted by the Fisher formalism. This serves as an internal consistency check of our Fisher Bias implementation: within the regime where the linear approximation underlying Equation \ref{eq:fisher_bias} is expected to hold, the recovered bias should reproduce the imposed input offset. Further, if a biased model prediction is generated within the same likelihood model used for the fiducial Fisher analysis, internal consistency checks such as those described above can be used as a preliminary diagnostic to determine whether the Fisher Bias formalism remains reliable or whether more sophisticated methods (e.g., direct likelihood sampling through MCMC) are required to accurately quantify the impact of parameter mis-specification.

As the imposed parameter shifts increase in magnitude, departures between the input and recovered biases naturally emerge, reflecting the breakdown of the linear expansion rather than a failure of the formalism itself. In particular, we observe that imposed shifts exceeding approximately $1\sigma$ in $w_0$ (Figure \ref{fig:fisher_bias_w0}) and $0.5\sigma$ in $w_a$ (Figure \ref{fig:fisher_bias_wa}) lead to deviations of more than 10\% between the input and predicted biases. At larger offsets, couplings between parameters become increasingly significant, such that a nominally one-dimensional perturbation induces correlated responses in other cosmological parameters.
\section{Conclusion}
\label{sec:conclusion}
The Vera C. Rubin Observatory will survey the entire visible southern sky every three nights for a period of ten years through the Legacy Survey of Space and Time (LSST), providing statistical power that was previously inaccessible by photometric large-scale structure surveys. Cosmological probes, such as \threept analyses, rely on accurate, precise models of the underlying statistical matter distribution and on robust analysis choices that mitigate the impact of known model uncertainties. As these investigations can span a large parameter space, fast, consistent methods and procedures for conducting model-robustness tests are essential to ensure accurate, unbiased inference of cosmological parameters.

In this work, we use \augur, an open-source Python package, to demonstrate and validate a Fisher forecasting framework developed within the LSST Dark Energy Science Collaboration (DESC). This framework connects the existing publicly available ecosystem of DESC tools (the Core Cosmology Library (\ccl), \firecrown, and \mcpcov) for the first time in a self-consistent pipeline. We first demonstrated several internal verification and consistency tests to achieve robust Fisher matrix configurations for LSST Y1- and Y10-like \threept analyses with respect to analysis hyperparameters, including numerical derivative step sizes and the sparseness of sampled redshift distributions, as defined in the DESC Science Requirements Document (SRD). After following our procedures to find a stable configuration, we investigated how potentially unstable configurations can produce unreliable Figures of Merit (FOMs) across derivative step sizes and non-linear matter power spectrum prescriptions.

Our results indicated that not all non-linear power spectra models respond identically to cosmological parameters for our setup, leading to FOMs that can differ by tens of percent at our stable configuration, i.e., the configuration corresponding to which the Fisher matrix is consistent across a range of modeling choices/hyperparameters and by orders of magnitude for unstable configurations. We generally found that our results are insensitive to the derivative method used; however, adaptive fits to non-differentiable likelihoods may encounter local extrema in the hyperparameters or oscillations that do not correspond to the true underlying degeneracy directions of the local likelihood. We assessed the value of (top-hat) bin-averaging in obtaining stable configurations for Fisher analyses, noting lower variability in the calculated FOMs across hyperparameter settings.

A broader challenge for future forecasting efforts is the increasing dimensionality of realistic cosmological analyses, where parameter spaces of $\mathcal{O}(100)$ nuisance and cosmological parameters are becoming standard. In this regime, finite-difference Fisher calculations become both computationally expensive and increasingly sensitive to numerical hyperparameters such as derivative step size and interpolation choices. Modern differentiable cosmology frameworks based on automatic differentiation, such as $\tt JAX$-based inference pipelines \citep{Campagne_2023, Piras_2023}, can offer a promising alternative by enabling stable derivative calculations without explicit finite differencing. While \augur currently interfaces with the existing DESC forecasting ecosystem built around finite-difference derivatives, the validation and numerical stability studies presented here provide an important benchmark for future Fisher forecasting approaches built on differentiable infrastructure.

Additionally, \augur results were compared to external tools, including the DESC SRD pipeline and a \cosmosis + \firecrown pipeline. We find the external pipelines to agree well within our established criteria: Fisher matrices and FOMs were nearly exact when compared to the \cosmosis + \firecrown pipeline, and the DEFOM agreed to 10\% and 2\% for our Y1- and Y10-like setups, respectively, when compared to the SRD pipeline. We also compare our results to direct likelihood sampling via \polychord nested sampling. We compared correlation matrices of all parameters between the nested sampling run and \augur output to account for naturally optimistic contours of Fisher analyses, finding the relative difference between the correlation matrices to be no greater than $0.04$, well within our proposed agreement criteria of $0.1$. 

We finally conducted internal consistency tests to assess the applicability of the Fisher Bias in systematic bias testing. We find \augur to be internally consistent when applying small 1D shifts in our model parameter space. We also illustrate the limitations of Fisher Bias analyses, showing that the locality of the approximations used breaks down for large parameter biases. We find simple tests like this have value in gauging whether more robust methods should be used to characterize the impact of systematic biases, particularly for cases where contaminated data vectors exist in the same model and parameter space as the evaluated Fisher matrix. 

% \FloatBarrier

\section*{Acknowledgments}
\label{sec:acknowledgments}

\textbf{General acknowledgments.}
PR is partially supported by the Department of Energy Cosmic Frontier program, grant DE--SC0010118.
SS is supported by the Alexander von-Humboldt foundation. 
NEC acknowledges support from the project ``A rising tide: Galaxy intrinsic alignments as a new probe of cosmology and galaxy evolution'' (with project number VI.Vidi.203.011) of the Talent programme Vidi which is (partly) financed by the Dutch Research Council (NWO).
AL acknowledges support from the Swedish National Space Agency (Rymdstyrelsen) under Career Grant Project Dnr 2024-00171 and from the research project grant `Understanding the Dynamic Universe' funded by the Knut and Alice Wallenberg Foundation under Dnr KAW 2018.0067.
CGG was supported by the Beecroft Trust and the C\'esar Nombela Research Talent Attraction grant from the Community of Madrid (Ref. 2025-T1/TEC-36302).
CG is funded by the MICINN project PID2022-141079NB-C32. IFAE is partially funded by the CERCA program of the Generalitat de Catalunya.
This document was prepared by LSST DESC using the resources of the Fermi National
Accelerator Laboratory (Fermilab), a U.S. Department of Energy, Office of Science, Office of High Energy
Physics HEP User Facility. Fermilab is managed by Fermi Forward Discovery Group, LLC, acting under
Contract No. 89243024CSC000002.
The DESC acknowledges ongoing support from the Institut National de 
Physique Nucl\'eaire et de Physique des Particules in France; the 
Science \& Technology Facilities Council in the United Kingdom; and the
Department of Energy and the LSST Discovery Alliance
in the United States.  DESC uses resources of the IN2P3 
Computing Center (CC-IN2P3--Lyon/Villeurbanne - France) funded by the 
Centre National de la Recherche Scientifique; the National Energy 
Research Scientific Computing Center, a DOE Office of Science User 
Facility supported by the Office of Science of the U.S.\ Department of
Energy under Contract No.\ DE-AC02-05CH11231; STFC DiRAC HPC Facilities, 
funded by UK BEIS National E-infrastructure capital grants; and the UK 
particle physics grid, supported by the GridPP Collaboration.  This 
work was performed in part under DOE Contract DE-AC02-76SF00515.
This paper has undergone internal review in the LSST Dark Energy Science Collaboration. 
We thank Nikolina Šarčević for discussions on the Fisher formalism and numerical derivative validation.
We are grateful to the internal reviewers, Dani Leonard and Tianqing Zhang, who provided helpful feedback and suggestions on the manuscript.

\textbf{Software acknowledgments.}
This work made extensive use of open-source scientific software, and we are grateful to the developers and maintainers of these tools.
We acknowledge the Python programming language\footnote{\url{https://www.python.org}} as the primary development environment.
Core numerical functionality was provided by \numpy\ \citep{numpy} and \scipy\ \citep{scipy}, with visualization handled using \matplotlib\ \citep{matplotlib}.
Posterior sampling, analysis, and visualization are handled within the library via \getdist\ \citep{getdist}, which are core dependencies used for sampling, post-processing, and inference workflows.
The analysis pipeline utilizes \numdifftools\footnote{\url{https://numdifftools.readthedocs.io}}  and \derivkit\ \citep{derivkit} for derivative-based inferences.
Theoretical predictions for the matter power spectrum and two-point observables are computed using \ccl\ \texttt{v3.3.0} \citep{Chisari_2019_CCL}.
Cosmological likelihood evaluations and inferences are made possible through \cosmosis \citep{Zuntz_2015} and \firecrown \texttt{v1.14.0}.
Nested sampling of the likelihood utilizes \polychord \citep{Handley_2015}. 
\augur covariance computations are enabled by \mcpcov \texttt{v0.5.1}. 
This analysis also used the initial data products and software from the DESC-SRD repository\footnote{\url{https://github.com/CosmoLike/DESC_SRD}}, namely \cosmolike \citep{Krause_2017}.

\textbf{Author contributions.}
PR: Analysis lead, \augur, \ccl, \firecrown, and \mcpcov development, major role in manuscript writing. 
SS: \augur development, weekly discussions, major role in manuscript writing.
JS: \augur development, validation, and manuscript writing.
NEC: Contributions to project scope, discussions, initial development, and manuscript crafting.
AL: Contributions to \augur and \firecrown interface development and forecasting discussions.
MP: \augur and \firecrown development, code review.
RP: \augur development.
HP: Initial development of the project.
BS: Initial development of the project and documentation.
AS: DESC Builder, \augur development, manuscript feedback.
SV: \firecrown development, manuscript feedback.
CGG: DESC Builder, \mcpcov development.
EG: DESC Builder, \augur design.
CG: DESC Builder, initial \augur testing, MCP co-convener.
CDL: DESC Builder, \ccl, \firecrown, and \mcpcov development, MCP co-convener, Internal Reviewer of this manuscript.
AM: DESC Builder, \firecrown development.
JN: DESC Builder, \ccl development.

\bibliographystyle{apsrev4-1}

% You should give the same name for your .bbl as your main .tex
% since it is a requirement for posting on ArXiv.
\bibliography{main}
 
\end{document}